\documentclass[aps,pra,twocolumn,floatfix,dblfloatfix,groupedaddress,footinbib,notitlepage,showpacs]{revtex4-2}
\usepackage{times,bm,bbm,bbold,amssymb,amsbsy,amsthm,amsmath,amsfonts,graphicx,graphics,color,xcolor,hyperref}
\hypersetup{colorlinks,linkcolor=blue,citecolor=blue,urlcolor=blue}

\newcommand{\ignore}[1]{}
\newtheorem{remark}{Remark}

\usepackage[T1]{fontenc} 
\usepackage{newtxtext,newtxmath} 

\DeclareFontFamily{OT1}{pzc}{}
\DeclareFontShape{OT1}{pzc}{m}{it}
              {<-> s * [1.25] pzcmi7t}{}
\DeclareMathAlphabet{\mathpzc}{OT1}{pzc}
                                 {m}{it}

\begin{document}
\title{Coherent dynamical control of quantum processes}

\author{V. Rezvani}
\email{vahid_rezvani@physics.sharif.edu}
\affiliation{Department of Physics, Sharif University of Technology, Tehran 14588, Iran}

\author{A. T. Rezakhani}
\email{rezakhani@sharif.edu}
\affiliation{Department of Physics, Sharif University of Technology, Tehran 14588, Iran}

\begin{abstract}
Manipulation of a quantum system requires the knowledge of how it evolves. To impose that the dynamics of a system becomes a particular target operation (for any preparation of the system), it may be more useful to have an equation of motion for the dynamics itself---rather than the state. Here we develop a Markovian master equation for the process matrix of an open system, which resembles the Lindblad Markovian master equation. We employ this equation to introduce a scheme for optimal local coherent process control at target times, and extend the Krotov technique to obtain optimal control. We illustrate utility of this framework through several quantum coherent control scenarios, such as optimal decoherence suppression, gate simulation, and passive control of the environment, in all of which we aim to simulate a given terminal process at a given final time.
\end{abstract}

\date{\today}
\maketitle

\section{Introduction}
\label{sec:intro}

In a real world a quantum system cannot be fully isolated from its surrounding reservoir or environment. Such system-environment couplings in general lead to a nonunitary description of the dynamics of the system \cite{breuer_theory_2002,book:rivas,book:Schaller,correlation-picture}. Consequently, useful quantum resources of an open system, such as quantum coherence and correlations, often diminish rapidly. To mitigate such adversarial effects, it seems necessary to employ ideas from quantum error correction \cite{nielsen_quantum_2010,book:QEC} and quantum control theory \cite{book:D'Alessandro, book:q.control, boscain_introduction_2020, Jirari_Optimal_Control}, such as quantum feedback control \cite{lloyd_coherent_2000}, decoherence-free subspaces and subsystems \cite{lidar_review_2012}, and dynamical decoupling \cite{viola_dynamical_1999, Khodjasteh, Viola-review}.  

If the system is initially uncorrelated with the environment, its dynamics can be faithfully described by quantum ``operations'' or ``channels'' (completely positive, trace-preserving linear maps), or equivalently by ``process matrices'' \cite{nielsen_quantum_2010}. These objects relate the instantaneous density matrix (i.e., state) of the system to its initial density matrix. Numerous methods, such as quantum process tomography, have been developed for characterization of process matrices \cite{nielsen_quantum_2010, emerson_symmetrized_2007, bendersky_selective_2008, mohseni_quantum-process_2008, mohseni_equation_2009}. In addition, under some assumptions, one can obtain master equations for dynamics of the state \cite{breuer_theory_2002,book:rivas,book:Schaller}. These master equations enable one to see how manipulation of the preparation or system Hamiltonian by external agents can affect the state of the system at any time. The ability to manipulate system dynamics has spurred quantum control applications \cite{Rabitz-0,Rabitz-1,Rabitz-2,peirce_optimal_1988, sundermann_extensions_1999, zhu_rapid_1998, zhu_rapidly_1998, bartana_laser_2001, carlini_time-optimal_2006, carlini_time-optimal_2007, carlini_time_2008, Gordon, QAB, Clausen, brody_quantum_2012, Suter-RMP, pang_optimal_2017, cavina_optimal_2018, cavina_variational_2018}. 

However, for some applications, it may still be more useful to have dynamical or master equations which describe the dynamics of the dynamics (the process matrix) rather than the dynamics of the state (the density matrix). A relevant example is a control scenario where one is interested to achieve a particular quantum operation in a physical system by applying suitable control fields. Since here the operation is of interest, a dynamical equation for how the associated process matrix evolves can provide more direct information about the target operation. Such equations can be particularly useful in dissipative control or reservoir engineering scenarios \cite{poyatos_quantum_1996, myatt_decoherence_2000, beige_quantum_2000, diehl_quantum_2008, verstraete_quantum_2009, lin_dissipative_2013, fedortchenko_finite-temperature_2014, reseng-1, Basilewitsch_2019, PRR-1}. 

In this paper, assuming Markovian evolution of the open system, we derive an equation of motion for the process matrix (for a precursor study see Ref. \cite{mohseni_equation_2009}). Next we use this equation to construct a fairly general scheme for \textit{optimal} control of the dynamics of open quantum systems, where the achieved operation is guaranteed to have the highest fidelity with the desired operation. We restrict ourselves to coherent control operations, where an external field is applied locally only on the system and modifies its Hamiltonian (assuming that the field does not modify the environment or the way it acts on the system). We use this scheme to study optimal coherent strategies for gate simulation, decoherence suppression, and passive control of the environment. Our optimization strategy is based on developing an extension of the Krotov method for quantum processes. In decoherence suppression, an optimal control field is applied to the open system to simulate a unitary evolution at a specified time. In quantum gate simulation, we show how a quantum gate can be simulated optimally when we are confined to coherent manipulation of the open system. This optimal control framework also allows us to force the environment to act as if it were another environment with different properties. For example, we show that how one can modify the system Hamiltonian such that at a specified time a dissipative environment looks like a depolarizing environment. We illustrate these scenarios for a practical model of a Rydberg ion under typical decoherence. 

The structure of this paper is as follows. In Sec. \ref{sec:CohCont} we derive our master equation for the process matrix. In addition, in this section we introduce the main components of the optimal control theory to manipulate instantaneously the dynamics of an open system. Next in Sec. \ref{sec:KrotovMethod} we solve this dynamical optimization problem based on a monotonically convergent algorithm, i.e., the Krotov algorithm. We focus on optimal coherent control of terminal processes in Sec. \ref{sec:Application} and apply this to three different scenarios. Section \ref{sec:conc} concludes the paper. The appendixes include some details and derivations. 

\section{COHERENT CONTROL OF QUANTUM PROCESSES} 
\label{sec:CohCont}

Consider an open quantum system $S$ of $N-$dimensional Hilbert space which interacts with its surrounding environment $B$ with a large number of degrees of freedom. The central problem in the optimal control theory is the dynamical manipulation of such a system to attain a given objective under some constraints \cite{boscain_introduction_2020}. For example, the objective can be dictating the dynamics of the system at a predetermined final (terminal) time $t_{\mathrm{f}}$ to become as much as possible similar to a given target dynamics. This can be achieved by manipulating the system and/or its environment. In this paper, we introduce a procedure for the dynamical manipulation of the system $S$ by applying the external fields only to the system.

\subsection{Dynamical variable: Process matrix}

In our dynamical control scheme, we work with an object referred to as the ``process matrix,'' which has a pivotal role in the dynamics of the open system. In the following we recall the definition of this object and some of its important properties \cite{nielsen_quantum_2010}.

Assume that the initial state of the total system (main system $+$ environment) is in a tensor-product form as $\varrho (t_{0})=\varrho_{S}(t_{0})\otimes\sum_{i}r_{i}\vert b_{i}\rangle\langle b_{i}\vert$, where the set of eigenvalues and eigenvectors of the initial state of the environment $B$
is denoted by $\left\lbrace r_{i},\vert b_{i}\rangle\right\rbrace $. Hence the time evolution of $S$ is described by a completely positive and trace-preserving linear map in a Kraus representation form as \citep{nielsen_quantum_2010}
\begin{equation}
\mathpzc{E}_{(t,t_{0})}\big(\circ\big)=\textstyle{\sum}_{\lambda ,\mu=1}^{N^{2}}\chi_{\lambda\mu}(t,t_{0})\, C_{\lambda}\circ C_{\mu}^{\dag}, 
\label{eq.1}
\end{equation}
where $\lbrace C_{\lambda}\rbrace_{\lambda=1}^{N^{2}}$ is a fixed orthonormal operator basis for the $N^{2}-$dimensional Liouville space of $S$, such that $\mathrm{Tr}[C_{\lambda}^{\dag}C_{\mu}]=\delta_{\mu\nu}$. In Eq. \eqref{eq.1}, the ``process matrix'' ${\chi }(t,t_{0})$ is defined as
\begin{align}
&\qquad\quad\chi(t,t_{0})=\mathpzc{B}^{\dagger}(t,t_{0}) \,\mathpzc{B}(t,t_{0}),\label{eq.2-2}
\\
&\mathpzc{B}_{(i,j),\mu}(t,t_{0})=\sqrt{r_{i}}\,\mathrm{Tr}[\langle b_{i}\vert U^{\dagger}(t,t_{0})\vert b_{j}\rangle\, C_{\mu}], 
\label{eq.2-3}
\end{align} 
where $U(t,t_{0})$ is the unitary operator generated by the total time-dependent Hamiltonian. Here we consider that the system is driven by an external control field $V_{\mathrm{field}}(t)$, which acts only on the system. Thus the total Hamiltonian becomes $H(t)=H_{S}+V_{\mathrm{field}}(t)+H_{B}+H_{\mathrm{int}}$, where $H_{S}$ ($H_{B}$) is the system (environment) Hamiltonian, and $H_{\mathrm{int}}$ denotes the system-environment interaction. In addition, in practical applications it is useful to consider that the external control $V_{\mathrm{field}}(t)$ can be adjusted by some knobs $\bm{\epsilon}(t)=\lbrace\epsilon_{m}(t) \rbrace$ as 
\begin{equation}
V_{\mathrm{field}}(t)= \bm{\epsilon}(t)\cdot\bm{H}= \textstyle{\sum_{m}} \epsilon_{m}(t)H_{m},
\end{equation}
where $H_{m}$ are some fixed control operators. 

The process matrix \eqref{eq.2-2} is a positive-semidefinite matrix and relates the initial state of the system to its states at next times. From Eq. \eqref{eq.2-2} one can see the process matrix contains all information about the dynamics of system $S$, and in this sense it is the dynamics itself. The trace-preserving property of the linear map $\mathpzc{E}_{(t,t_{0})}$ at all times $t\geqslant t_{0}$ implies that $\mathrm{Tr}[\chi (t,t_{0})]=N$. There is a one-to-one isomorphism, refereed to as the Choi-Jamiolkowski isomorphism \cite{Jamiolkowski, Choi}, between any completely positive map $\mathpzc{E}_{(t,t_{0})}$ and the density matrix of a composite system comprised of the open system $S$ and an ancilla of the same Hilbert space dimension, which is given by $\varrho_{\mathpzc{E}}(t)=\big(\mathpzc{E}_{(t,t_{0})}\otimes\mathbbmss{I}_{S}\big)\big(\vert\Phi_{+}\rangle\langle\Phi_{+}\vert\big)$, where $\vert\Phi_{+}\rangle$ is a maximally entangled state of the composite system. One can see that in the logical operator basis $\lbrace\widetilde{C}_{(i,j)}=\vert i\rangle\langle j\vert\rbrace_{i,j=1}^{N}$, with $\vert i\rangle$ being the computational basis, the process matrix $\widetilde{\chi}(t,t_{0})$ is proportional to the corresponding density matrix $\varrho_{\mathpzc{E}}(t)$ \cite{gilchrist_distance_2005}, 
\begin{equation}
\widetilde{\chi} (t,t_{0})=N\varrho_{\mathpzc{E}}(t).
\label{eq.2-1}
\end{equation}
In addition, the process matrix $\widetilde{\chi}(t,t_{0})$ is related to process matrix ${\chi}(t,t_{0})$ in an arbitrary operator basis $\{ C_{\alpha}\}$ through a unitary transformation $\widetilde{\chi}(t,t_{0})=\mathcal{S}^{\dag}{\chi}(t,t_{0})\mathcal{S}$, where the unitary operator $\mathpzc{S}$ is defined as $\mathcal{S}_{\alpha ,(i,j)}=\mathrm{Tr}[C_{\alpha}^{\dag}\widetilde{C}_{(i,j)}]$ for $\alpha\in\{1,\ldots,N^{2}\}$ and $i,j\in\{1,\ldots,N\}$.

\subsection{Dynamical equation of the dynamics}

After defining the process matrix as a dynamical variable, we need to determine how the process matrix evolves by considering the dissipative and external field effects in the dynamics space $\mathfrak{D}_{S} =\left\lbrace \chi (t,t_{0});\,\forall t\geqslant t_{0}\right\rbrace$.

Here we restrict ourself to the special case of quantum Markovian evolutions \citep{book:rivas}, where the dynamical map \eqref{eq.1} satisfies the divisibility condition for all $t,s$ such that $t_{0}\leqslant s\leqslant t$,
\begin{equation}
\mathpzc{E}_{(t,t_{0})}=\mathpzc{E}_{(t,s)}\mathpzc{E}_{(s,t_{0})}.
\label{eq.3-1}
\end{equation}
It is straightforward to see that this condition on the dynamics in turn implies the following Markovian master equation in the Lindblad form for the density matrix $\varrho_{S}(t)$:
\begin{align}
\dfrac{d\varrho_{S}(t)}{dt}=&-\dfrac{i}{\hbar}[{H}_{S}(t),\varrho_{S}(t) ]+\textstyle{\sum}_{\alpha=1}^{N^{2}-1}\gamma_{\alpha}(t) 
\big(L_{\alpha}(t)\varrho_{S}(t) L_{\alpha}^{\dag}(t)
\nonumber\\
&-\dfrac{1}{2}\lbrace L_{\alpha}^{\dag}(t)L_{\alpha}(t),\varrho_{S}(t)\rbrace \big),
\label{eq.3-2}
\end{align}
where the Lindblad operators are defined as $L_{\alpha}(t)=\sum_{\eta =1}^{N^{2}-1}u_{\alpha\eta}^{\ast}(t) C_{\eta}$, with $u(t)$ being the unitary operator diagonalizing the positive semidefinite matrix $a(t)=[a_{\xi\nu}(t)]=[\lim_{x\rightarrow 0}\chi_{\xi\nu}(t+x,t)/x]$ (for $\xi,\nu\in\{1,\ldots, N^{2}-1\}$) as $u(t)a(t)u^{\dag}(t)=\mathrm{diag}\big(\gamma_{\alpha}(t) \big)$. The coefficients $\gamma_{\alpha}(t)$ are referred to as the Lindblad rates. The Hermitian operator $H_{S}(t)$ is also defined as $H_{S}(t)=\big(M(t)-M^{\dag}(t) \big)/(2i)$, where $M(t)=(\hbar /\sqrt{N})\sum_{\lambda=1}^{N^{2}-1}a_{\lambda N^{2}}(t)C_{\lambda}$, $a_{\lambda N^{2}} (t)=\lim_{x\rightarrow 0}\chi_{\lambda N^{2}}(t+x,t)/x$. 

Interestingly an alternative microscopic, first-principle approach to derive the above Lindblad master equation has also been formulated \cite{breuer_theory_2002}, which can help to find the validity conditions for Eq. \eqref{eq.3-2} and also underlying physical meanings of its various components such as $H_{S}(t)$. This approach is based on the dynamics of the total system described by the von Neumann equation $\frac{d}{dt}\varrho_{SB}(t)=-(i/\hbar)[H(t),\varrho_{SB}(t)]$, partial tracing over the environment, and the weak-coupling, Born-Markov, and the secular approximations \cite{breuer_theory_2002,book:Schaller, lidar2019lecture}. The weak-coupling approximation implies that 
\begin{equation}
\Vert H_{\mathrm{int}}\Vert \ll \max_{t}\Vert H_{S}+V_{\mathrm{field}}(t)+H_{B}\Vert,
\label{eq.3-3}
\end{equation}
where $\Vert\cdot\Vert$ is the standard operator norm. In addition, Eq. (\ref{eq.3-2}) requires validity of particular assumptions about several time scales in the total system. In particular, 
\begin{gather}
\tau_{B} \ll \delta t_{S}\ll\tau_{S}, \label{cond-1}\\
1/\min_{\omega\neq \omega'}|\omega-\omega'|\ll \delta t_{S}, 
\label{cond-2}
\end{gather}
where $\tau_{B}$ is the relaxation time of the environment (the time at which the correlation functions of the environment decay), $\delta t_{S}\approx 1/\max_{t}\Vert H_{S}+V_{\mathrm{field}}(t)\Vert$ is the time scale of the variations of the driven system, $\tau_{S}$ is the time scale for the relaxation of the systems, and $\omega$ (or $\omega'$) indicates the energy gaps of $H_{S}$. The assumptions \eqref{eq.3-3} -- \eqref{cond-2} lead to time-independence of the Lindblad rates and operators \cite{Geva,Tannor-1}. In some sense, these assumptions guarantee that the control fields do not considerably modify how the environment affects the system. In addition, through the microscopic approach, the Hermitian operator $H_{S}(t)$ in Eq. \eqref{eq.3-2} appears to be the sum of the bare system, the field-system interaction, and the time-independent Lamb-shift Hamiltonian (environment-induced correction to the bare system Hamiltonian), $H_{S}(t)=H_{S}+V_{\mathrm{field}}(t)+H_{\mathrm{Lamb}}$. This form is in the \textit{lab} frame.

To obtain a master equation for the dynamics $\chi (t,t_{0})$ in the Markovian regime, first we translate the divisibility condition \eqref{eq.3-1} in the language of the process matrix,
\begin{equation}
 \chi_{\alpha\beta}(t,t_{0})=\textstyle{\sum}_{\mu,\eta=1}^{N^{2}}\textstyle{\sum}_{\lambda,\nu=1}^{N^{2}}\chi_{\lambda\nu}(t,s)\chi_{\mu\eta}(s,t_{0})\,[ F_{\lambda}]_{\alpha\mu} \,[ F_{\nu}]^{*}_{\beta\eta},
\label{eq.3}
\end{equation}
where $F$ is a rank-$3$ tensor defined as $[F_{\lambda}]_{\alpha\mu}\equiv\mathrm{Tr}[C_{\alpha}^{\dag}C_{\lambda}C_{\mu}]$. From Eq. \eqref{eq.3} and by differentiating the process matrix, we can obtain a linear differential equation as
\begin{align}
\dfrac{d\chi(t,t_{0})}{dt}=-\dfrac{i}{\hbar} \mathpzc{K}\big(\chi(t,t_{0})\big), \label{eq.4}
\end{align}
for $\alpha,\beta\in\{1,\ldots,N^{2}\}$, where the time-dependent generator $\mathpzc{K}$ is given by
\begin{align}
-\dfrac{i}{\hbar}[\mathpzc{K}]_{\alpha\beta,\mu\eta}=&\lim_{x\rightarrow 0}\frac{1}{x}\big(\textstyle{\sum}_{\lambda,\nu=1}^{N^{2}}\chi_{\lambda\nu}(t+x,t) [ F_{\lambda}]_{\alpha\mu}
 [ F_{\nu}]^{*}_{\beta\eta}
\nonumber \\
&-\delta_{\alpha\mu} \delta_{\beta\eta}\big).
\label{eq.5}
\end{align}
Following the next steps similar to the formal derivation of the Lindblad equation for the density matrix yields an expression for the generator $\mathpzc{K}$. We have relegated the details to Appendix \ref{app:a}. From hereon we assume $t_{0}=0$ and use the shorthand $(t)$ for $(t,0)$. In addition, we assume the basis operators $\left\lbrace C_{\alpha}\right\rbrace$ are traceless, $\mathrm{Tr}[C_{\alpha}]=0$ for $\alpha\in\{1,\ldots, N^{2}-1\}$, except for $C_{N^{2}}=\mathbbmss{I}_{S}/\sqrt{N}$. These steps transform the dynamical master equation of the system (\ref{eq.3-2}) into the following Markovian dynamical equation for the process matrix itself:
\begin{align}
\dfrac{d\chi (t)}{dt}=&-\frac{i}{\hbar}\mathpzc{K}\big(\chi(t)\big) 
=-\dfrac{i}{\hbar} [\mathsf{H}_{S}(t),\chi(t) ]+\textstyle{\sum}_{\alpha=1}^{N^{2}-1}\gamma_{\alpha}(t) \nonumber\\
&\times\big(\mathsf{L}_{\alpha}(t)\chi(t) \mathsf{L}_{\alpha}^{\dag}(t)
-\dfrac{1}{2}\lbrace \mathsf{L}_{\alpha}^{\dag}(t)\mathsf{L}_{\alpha}(t),\chi(t)\rbrace \big),
\label{eq.6}
\end{align}
where $\gamma_{\alpha}(t)$'s are the same factors as in the Lindblad equation \eqref{eq.3-2}, 
\begin{equation}
[\mathsf{Y}(t)]_{\mu\nu}\equiv\mathrm{Tr}[C_{\mu}^{\dag}\,Y(t)\, C_{\nu}],
\label{eq.7}
\end{equation}
with $Y(t)\in\{H_{S}(t), L_{1}(t),\ldots,L_{N^{2}-1}(t)\}$, and the initial value condition is $\chi_{\mu\nu}(0)=N\delta_{\mu N^{2}}\delta_{\nu N^{2}}$, for $\mu,\nu\in\{1,\ldots, N^{2}\}$. Equation (\ref{eq.6}), which describes the dynamics, is one of the main results of this paper, and has evident similarity with Eq. (\ref{eq.3-2}). It is straightforward to see that the conditions \eqref{eq.3-3} -- \eqref{cond-2} yield that $\gamma_{\alpha}(t)$'s and $\mathsf{L}_{\alpha}(t)$'s become time-independent. In addition, for simplicity and noting that the Lamb-shift correction is of the second-order with respect to $H_{\mathrm{int}}$, we neglect $H_{\mathrm{Lamb}}$. 

It is important to note that the above equation directly addresses the dynamics of an open system without reference to its state. This is an appealing property which can be of significant practical importance, especially when one is interested to design or simulate a particular operation, dynamics, or gate, rather than a particular state. In this sense, our dynamical equation (\ref{eq.6}) can be taken as the basis for an enhanced quantum dynamical control scheme for open systems, which may have numerous applications in diverse areas, such as quantum computation \cite{koch_controlling_2016}. This lifting from the dynamics of a state to the dynamics of the dynamics can be compared with the closed-system scenario in which rather than the Schr\"{o}dinger equation $\frac{d}{dt}\varrho_{S}(t)=-(i/\hbar)[H_{S}(t),\varrho_{S}(t)]$ one can work with the dynamical equation $\frac{d}{dt}U_{S}(t,0)=-(i/\hbar)H_{S}(t)\,U_{S}(t,0)$, where $\varrho_{S}(t)=U_{S}(t,0)\,\varrho_{S}(0)\,U^{\dag}_{S}(t,0)$.

\begin{remark}
From Eqs. \eqref{eq.2-1} and \eqref{eq.6} one can obtain an equation for the Choi-Jamiolkowski density matrix $\varrho_{\mathpzc{E}}(t)$ as
\begin{align}
\dfrac{d\varrho_{\mathpzc{E}}(t)}{dt}=&-\dfrac{i}{\hbar}[\widetilde{\mathsf{H}}_{S}(t),{\varrho_{\mathpzc{E}}}(t) ]+\textstyle{\sum}_{\alpha=1}^{N^{2}-1}\gamma_{\alpha}(t)  \big(\widetilde{\mathsf{L}}_{\alpha}(t){\varrho_{\mathpzc{E}}}(t) \widetilde{\mathsf{L}}_{\alpha}^{\dag}(t)
\nonumber\\
&-\dfrac{1}{2}\lbrace \widetilde{\mathsf{L}}_{\alpha}^{\dag}(t) \widetilde{\mathsf{L}}_{\alpha}(t) , {\varrho_{\mathpzc{E}}}(t)\, \rbrace \big),
\label{eq.9-1}
\end{align}
where $\widetilde{\mathsf{Z}}(t)$ is defined as $\widetilde{\mathsf{Z}}(t)=\mathcal{S}^{\dag}\mathsf{Z}(t)\mathcal{S}$, with $\mathsf{Z}(t)\in\{\mathsf{H}_{S}(t), \mathsf{L}_{1}(t),\ldots,\mathsf{L}_{N^{2}-1}(t)\}$. 
\end{remark}

\subsection{Objective of the control}

To analyze how effectively the applied fields $\bm{\epsilon}(t)$ performs toward achieving our objective, we need to choose a relevant figure-of-merit. A general figure-of-merit to control dynamics of the open system $S$ is a real scalar functional in the form of
\begin{equation}
\mathpzc{J}=\mathpzc{F}\big(\chi(t_{\mathrm{f}}),t_{\mathrm{f}}\big)+\mathpzc{J}_{d}[\chi(t)]+\mathpzc{J}_{f}[\bm{\epsilon}(t)].
\label{eq.10}
\end{equation}
The final time-dependent objective $\mathpzc{F}$ can be constructed based on a measure which compares how close the achieved dynamics at $t=t_{\mathrm{f}}$, $\chi(t_{\mathrm{f}})$, is to a given desired dynamics, $\Xi_{\,t_{\mathrm{f}}}$. For example, we can employ the quantum operator fidelity defined as \cite{Wang-fidelity,Suter-RMP}
\begin{equation}
\mathpzc{F}\big(\chi(t_{\mathrm{f}}),t_{\mathrm{f}}\big)=- w_{0}\dfrac{\mathrm{Tr}[\chi^{\dag}(t_{\mathrm{f}})\, \Xi_{\,t_{\mathrm{f}}}]}{\big(\mathrm{Tr}[\chi^{\dag}(t_{\mathrm{f}})\chi(t_{\mathrm{f}})]\,\mathrm{Tr}[\Xi^{\dag}_{\,t_{\mathrm{f}}}\Xi_{\,t_{\mathrm{f}}}]\big)^{1/2}},
\label{eq.11}
\end{equation}
where $ w_{0}\geqslant 0$ is a weight and the negative sign is a convention put to simply transform the optimization into a minimization problem. This quantity is guaranteed by the Cauchy-Schwarz inequality to be bounded as $- w_{0}\leqslant \mathpzc{F}\big(\chi(t_{\mathrm{f}}),t_{\mathrm{f}}\big)\leqslant 0$. For simplicity, later in this paper, when we discuss several examples in Sec. \ref{sec:Application}, we shall restrict ourselves to the case $w_{0}=1$, in which case $-\mathpzc{F}$ becomes the process fidelity.

Note that in the context of optimal open-system quantum control theory, to achieve a desired quantum gate $O$ at a predetermined terminal time $t_{\mathrm{f}}$, various measures or fidelities have been proposed to represent the final time-dependent objective $\mathpzc{F}$ \cite{koch_controlling_2016}. However, majority of these measures are based on carving the dynamics on pure or logical states, rather than taking the dynamics into account directly. For example, one can name the pure state-based fidelity \cite{nielsen_simple_2002, goerz_robustness_2014} such as $\mathpzc{F}_{\mathrm{p}}=\textstyle{\int}d\vert\Psi\rangle\,\langle\Psi\vert O^{\dag}\mathpzc{E}_{(t_{\mathrm{f}},0)}\big(\vert \Psi\rangle\langle\Psi\vert\big)O\vert\Psi\rangle$ and the logical basis-based one \cite{goerz_optimal_2014, koch_controlling_2016} such as $\mathpzc{F}_{l}=(1/N)\sum_{i,j=1}^{N}\mathrm{Tr}[O\vert i\rangle \langle j\vert O^{\dag}\mathpzc{E}_{(t_{\mathrm{f}},0)}(\vert i\rangle\langle j\vert)]$. In contrast, our dynamics-based objective \eqref{eq.11} is directly related to the dynamics, without reference to any state.

The functional $\mathpzc{J}_{d}$ depends on the dynamics of the system, $\chi (t)$, at the intermediate times $0\leqslant t< t_{\mathrm{f}}$ . In order to steer the dynamics $\chi (t)$ toward a desired one $\Xi (t)$ through the external fields, one can write this intermediate time-dependent objective $\mathpzc{J}_{d}$ as
\begin{align}
\mathpzc{J}_{d}[\chi(t)]=-\dfrac{ w_{d}}{t_{\mathrm{f}}} \int_{0}^{t_{\mathrm{f}}}dt\;\dfrac{\mathrm{Tr}[\chi^{\dag}(t)\, \Xi(t)]}{\big(\mathrm{Tr}[\chi^{2}(t)]\,\mathrm{Tr}[\Xi^{2}(t)]\big)^{1/2}},
\label{eq.12}
\end{align}
where $w_{d}\geqslant 0$ is a weight for this objective. In this work, however, in order to demonstrate basic utility of our proposed control scheme, we restrict ourselves to the dynamical control problem of the system at a given terminal time $t_{\mathrm{f}}$, rather than at an interval, which means that we simply set $w_{d}= 0$ in our study. Nevertheless, our scheme can be straightforwardly applied to the cases where one aims to control a dynamics $\chi(t)$ so that it can resemble a desired dynamics $\Xi(t)$ for a given time interval $0\leqslant t< t_{\mathrm{f}}$.
\begin{figure}
\includegraphics[scale=0.38]{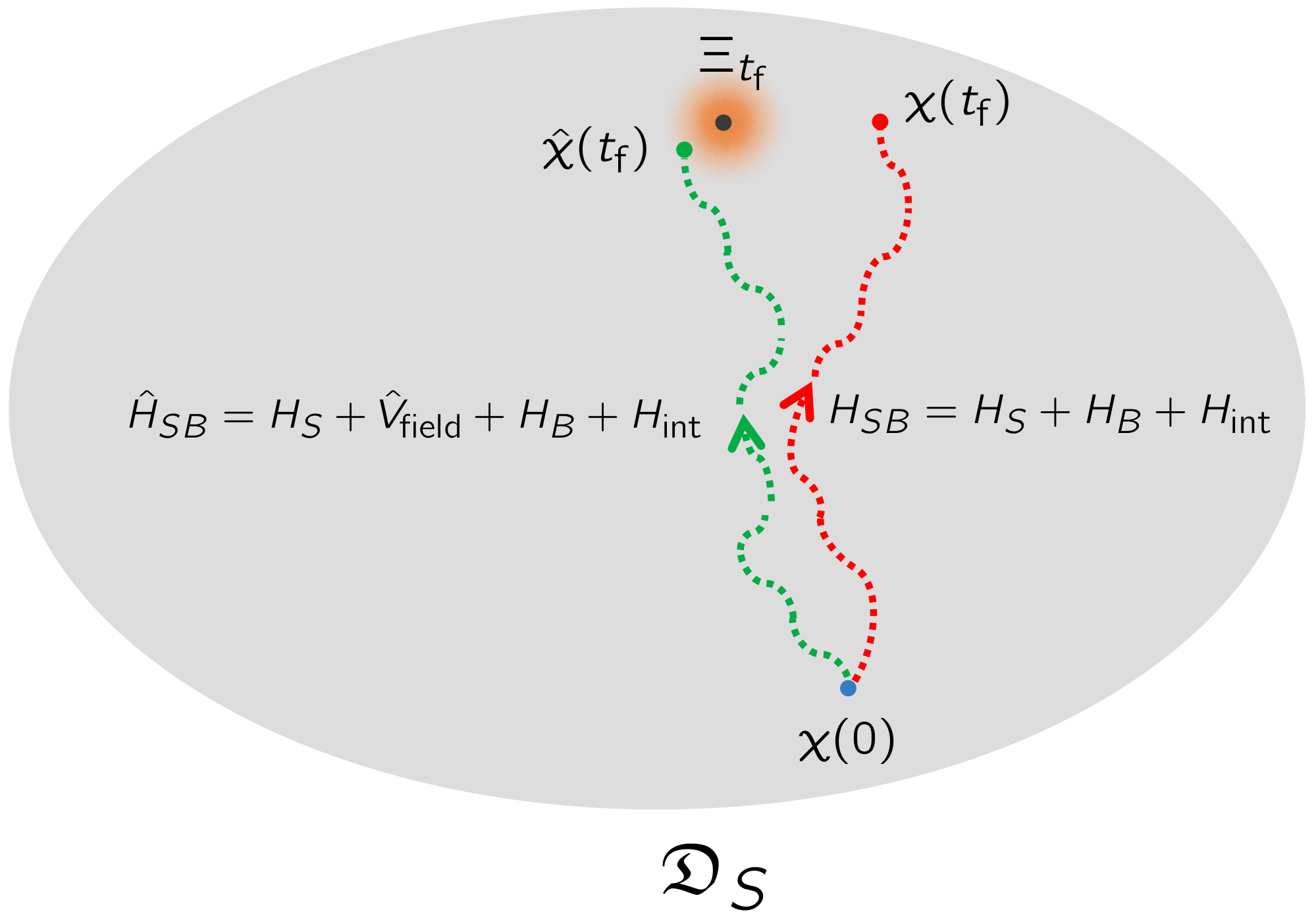}
\caption{Schematic of the optimal control problem for the terminal process. This dynamical control occurs in the space $\mathfrak{D}_{S}$ containing all process matrices. The red dotted curve shows the trajectory of the dynamics of an open quantum system $S$ generated by $H_{SB}=H_{S}+H_{B}+H_{\mathrm{int}}$. The green dashed curve indicates the trajectory of the dynamics generated by the optimal Hamiltonian $\hat{H}_{SB}(t)=H_{SB}+\hat{V}_{\mathrm{field}}(t)$. This Hamiltonian forces the terminal system process at $t=t_{\mathrm{f}}$ to be optimally close to the target terminal process $\Xi_{\,t_{\mathrm{f}}}$.}
\label{Fig-1}
\end{figure}

A clear advantage of our objective measures \eqref{eq.11} and \eqref{eq.12} is that, in light of the existence of various experimental techniques such as process tomography to estimate process matrices \cite{mohseni_quantum-process_2008}, they are also experimentally accessible. 

The field-dependent functional $\mathpzc{J}_{f}$ has been introduced to take into account all operational or experimental constraints on the control fields. For example, we assume
\begin{gather}
\mathpzc{J}_{f} [\bm{\epsilon}(t)]=\textstyle{\int_{0}^{t_{\mathrm{f}}}}dt\,\mathcal{G}_{f}\big(\bm{\epsilon}(t),t\big),\label{eq.13.1} \\
\mathcal{G}_{f}\big(\bm{\epsilon}(t),t\big)=\textstyle{\sum}_{m} w_{m} \big(\epsilon_{m}(t)-\epsilon_{m}^{(\mathrm{ref})}(t) \big)^{2}/f_{m}(t),
\label{eq.13}
\end{gather}
where $w_{m}\geqslant 0$ and $\epsilon_{m}^{(\mathrm{ref})}(t)$ are, respectively, a weight for this objective and a reference field. The shape function $f_{m}(t)$ can switch the external field $\epsilon_{m}(t)$ on and off smoothly \cite{sundermann_extensions_1999} (see also Refs. \cite{Gordon,Clausen} for energy-constrained protocls). By setting $\epsilon_{m}^{(\mathrm{ref})}(t)=0$, the functional \eqref{eq.13.1} implies a constraint on the energy of the control field.

The main goal of our dynamical control scheme is to minimize the total functional $\mathpzc{J}$ \eqref{eq.10} containing only the final time- and the field-dependent objectives [Eqs. \eqref{eq.11} and \eqref{eq.13.1}] such that the dynamics of the system is governed by the master equation \eqref{eq.6}. Figure \ref{Fig-1} depicts a schematic of this scheme. In the following section, we proceed to solve this problem via a monotonically convergent optimization approach; the Krotov method \cite{book:Krotov, Krotov_method, sklarz_loading_2002,Schirmer-Krotov,Morzhin-Pechen,Goerz_2019}.

\section{Solving the dynamical optimization problem: Krotov method}
\label{sec:KrotovMethod}

To solve the optimal control problem discussed in the previous section, we can resort to one of existing iterative methods. Here in particular we focus on monotonically convergent methods. One of such methods is the Zhu-Rabitz (ZR) method \cite{zhu_rapid_1998, zhu_rapidly_1998}, where after obtaining the control equations by variational approaches, a particular monotonically convergent algorithm is employed to solve these equations. In the generalized form of this technique \cite{maday_new_2003, ohtsuki_generalized_2004}, the speed of convergence and numerical accuracy of the algorithm can also be adjusted via appropriate convergence parameters. An alternative method to solve an optimization problem in a monotonically convergent fashion is the Krotov method \cite{Krotov_method, sklarz_loading_2002}. Having a dynamical equation and an objective functional, in this method a particular algorithm is employed to improve the control objective after each iteration. This method has been used in controlling states of closed or open systems, and it encompasses a wide range of optimization problems including nonconvex state-dependent functionals, intermediate time-dependent functionals, and nonlinear master equation for the state \cite{sklarz_loading_2002, reich_monotonically_2012}. 

Note that the ZR method is reduced to the Krotov method by choosing the convergence parameters suitably \cite{maday_new_2003}. Although for some convergence parameters the ZR method exhibits faster convergence and also a higher numerical accuracy than the Krotov method \cite{ohtsuki_generalized_2004}, to the best of our knowledge the ZR method has not yet been generalized to the case of nonconvex final time-dependent objectives such as the one we have introduced in Eq. \eqref{eq.11}. Hence here we focus on the Krotov method and extend it to the problem of controlling the process of an open system. We have relegated the details to Appendix \ref{app:b} and only mention the main results here in the following.

In the following, it seems more convenient to vectorize the dynamics (process matrix) $\chi (t)$ and the field-dependent generator (denoted by $\mathpzc{K}_{\,\,\bm{\epsilon}}$) \cite{ohtsuki_bathinduced_1989, metrology_us} in an extended Hilbert space as $\chi(t)\, \rightarrow\vert\chi (t)\rangle\hskip-0.7mm\rangle$ and $\mathpzc{K}_{\,\,\bm{\epsilon}}\rightarrow \mathbbmss{K}_{\bm{\epsilon}}$, respectively. For any $\vert\chi_{1}\rangle\hskip-0.7mm\rangle$ and $\vert\chi_{2}\rangle\hskip-0.7mm\rangle$ in this space, the inner product is defined as $\langle\hskip-0.7mm\langle\chi_{1}\vert\chi_{2}\rangle\hskip-0.7mm\rangle =\mathrm{Tr}[\chi_{1}^{\dag}\chi_{2}]$. In this representation the final time-dependent objective \eqref{eq.11} takes the following form:
\begin{equation}
\mathpzc{F}\big(\chi(t_{\mathrm{f}}),t_{\mathrm{f}}\big)=- w_{0}\dfrac{\langle\hskip-0.7mm\langle\chi(t_{\mathrm{f}})\vert\Xi_{\,t_{\mathrm{f}}}\rangle\hskip-0.7mm\rangle}{[\langle\hskip-0.7mm\langle\chi(t_{\mathrm{f}})\vert\chi(t_{\mathrm{f}})\rangle\hskip-0.7mm\rangle\langle\hskip-0.7mm\langle\Xi_{\,t_{\mathrm{f}}}\vert\Xi_{\,t_{\mathrm{f}}}\rangle\hskip-0.7mm\rangle]^{1/2}}.
\label{eq.14}
\end{equation}
Each iteration of the Krotov method contains two consecutive steps. At the first step, we have the dynamics $\vert \chi^{(n)} (t)\rangle\hskip-0.7mm\rangle$ as an input of the current iteration $(n+1)$ ($n\geqslant 0$) governed by the dynamical equation $d\vert \chi^{(n)} (t)\rangle\hskip-0.7mm\rangle /dt=-(i/\hbar)\mathbbmss{K}_{\bm{\epsilon}^{(n)}}\vert\chi ^{(n)}(t)\rangle\hskip-0.7mm\rangle$, with the initial boundary condition on the $\alpha$'s component $\vert \chi^{(n)}(0)\rangle\hskip-0.7mm\rangle_{\alpha} =N\delta_{\alpha N^{4}}$, where $\mathbbmss{K}_{\bm{\epsilon}^{(n)}}$ shows the time-dependent generator with $\bm{\epsilon}^{(n)}(t)$. The $\bm{\epsilon}^{(n)}(t)$ are the control fields updated in the previous iteration (and when $n=0$, these are the initial guess fields). Having these fields, an adjoint dynamics $\vert \Lambda (t) \rangle\hskip-0.7mm\rangle$ evolves backward in time according to
\begin{equation}
\dfrac{d\vert \Lambda (t) \rangle\hskip-0.7mm\rangle}{dt}=-\dfrac{i}{\hbar}\mathbbmss{K}^{{\dagger}}_{\bm{\epsilon}^{{(n)}}}\vert\Lambda (t) \rangle\hskip-0.7mm\rangle,
\label{eq.15}
\end{equation}
with the following boundary condition at the final time $t_{\mathrm{f}}$:
\begin{equation}
\vert \Lambda (t_{\mathrm{f}})\rangle\hskip-0.7mm\rangle =-\Big( \dfrac{\partial\mathpzc{F}}{\partial\langle\hskip-0.7mm\langle\chi (t_{\mathrm{f}})\vert}\Big)\Big| _{\chi^{(n)}(t_{\mathrm{f}})},
\label{eq.16}
\end{equation}
where $|_{\chi^{(n)}(t_{\mathrm{f}})}$ on the right-hand side (RHS) indicates the evaluation of the partial derivative at $\vert\chi ^{(n)}(t_{\mathrm{f}})\rangle\hskip-0.7mm\rangle$ and $\langle\hskip-0.7mm\langle\chi ^{(n)}(t_{\mathrm{f}})\vert$. Note that for the full control of an open system at a specified time interval---that is, when $\mathpzc{J}_{d}\neq 0$---the RHS of Eq. \eqref{eq.15} needs to be modified \cite{reich_monotonically_2012}. For the dynamical control of a system at a predetermined time with the final time-dependent objective \eqref{eq.14}, the boundary condition \eqref{eq.16} can be written as
\begin{align}
\vert \Lambda (t_{\mathrm{f}})\rangle\hskip-0.7mm\rangle =&\dfrac{ w_{0}}{2}\Big(\dfrac{\vert {\Xi}_{t_{\mathrm{f}}}\rangle\hskip-0.7mm\rangle}{\big[\langle\hskip-0.7mm\langle \chi^{(n)}(t_{\mathrm{f}})\vert \chi^{(n)}(t_{\mathrm{f}})\rangle\hskip-0.7mm\rangle\langle\hskip-0.7mm\langle\Xi_{\,t_{\mathrm{f}}}\vert\Xi_{\,t_{\mathrm{f}}}\rangle\hskip-0.7mm\rangle\big]^{1/2}}
\nonumber\\
&-\dfrac{\langle\hskip-0.7mm\langle \chi^{(n)}(t_{\mathrm{f}})\vert {\Xi}_{t_{\mathrm{f}}}\rangle\hskip-0.7mm\rangle\vert\chi^{(n)}(t_{\mathrm{f}})\rangle\hskip-0.7mm\rangle}{\big[\langle\hskip-0.7mm\langle\chi^{(n)}(t_{\mathrm{f}})\vert\chi^{(n)}(t_{\mathrm{f}})\rangle\hskip-0.7mm\rangle^{3}\langle\hskip-0.7mm\langle\Xi_{\,t_{\mathrm{f}}}\vert\Xi_{\,t_{\mathrm{f}}}\rangle\hskip-0.7mm\rangle\big]^{1/2}}\Big).
\label{eq.17}
\end{align}

At the next step of the \textcolor{blue}{$n+1$}th iteration, the control field $\epsilon_{m}(t)$ needs to be updated according to the equation
\begin{align}
&\Big(\dfrac{\partial\mathcal{G}_{f}}{\partial\epsilon_{m}}\Big)\Big|_{\bm{\epsilon}^{(n+1)}}=\dfrac{2}{\hbar}\mathrm{Im}\Big\{\langle\hskip-0.7mm\langle\Lambda(t)\, \vert\Big(\dfrac{\partial\mathbbmss{K}_{\bm{\epsilon}}}{\partial\epsilon_{m}}\Big)\Big| _{\bm{\epsilon}^{(n+1)}}\vert\chi^{(n+1)}(t)\rangle\hskip-0.7mm\rangle\Big\}
\nonumber\\
&\,+\dfrac{\sigma(t)}{\hbar}\mathrm{Im}\Big\{\langle\hskip-0.7mm\langle\Delta\chi^{(n+1)}(t)\, \vert\Big(\dfrac{\partial\mathbbmss{K}_{\bm{\epsilon}}}{\partial\epsilon_{m}}\Big)\Big| _{\bm{\epsilon}^{(n+1)}}\vert\chi^{(n+1)}(t)\rangle\hskip-0.7mm\rangle\Big\},
\label{eq.18}
\end{align}
where $\vert \Delta\chi^{(n+1)}(t)\rangle\hskip-0.7mm\rangle=\vert \chi^{(n+1)}(t)\rangle\hskip-0.7mm\rangle -\vert \chi^{(n)}(t)\rangle\hskip-0.7mm\rangle$ is the change in the dynamics and $\vert \chi^{(n+1)}(t)\rangle\hskip-0.7mm\rangle$ follows the dynamical equation $d\vert \chi^{(n+1)} (t)\rangle\hskip-0.7mm\rangle /dt=-(i/\hbar)\mathbbmss{K}_{\bm{\epsilon}^{(n+1)}}\vert\chi ^{(n+1)}(t)\rangle\hskip-0.7mm\rangle$ with $\mathbbmss{K}_{\bm{\epsilon}^{(n+1)}}$. These two equations are coupled and should be solved simultaneously. Due to the nonconvexity of the final time-dependent objective \eqref{eq.14}, a coefficient $\sigma(t)$ has been introduced in Eq. \eqref{eq.18} to guarantee monotonic convergence of the ultimate algorithm. This time-dependent coefficient is determined analytically by the relation
\begin{equation}
\sigma(t)=-\bar{A}e^{\zeta_{B}(t_{\mathrm{f}}-t)},\qquad\zeta_{B}\in\mathbbmss{R}^{+}, 
\label{eq.19.1}
\end{equation}
where the constant coefficient $\bar{A}$ is given by
\begin{gather}
\bar{A}=\mathrm{max}\{\zeta_{A},2A+\zeta_{A}\},\qquad\zeta_{A}\in\mathbbmss{R}^{+},
\label{eq.19.2.1} \\
A=\sup_{\{\Delta\chi(t_{\mathrm{f}})\}}\dfrac{\Delta\mathpzc{F}+2\mathrm{Re}\langle\hskip-0.7mm\langle\Delta\chi(t_{\mathrm{f}})\vert\Lambda(t_{\mathrm{f}})\rangle\hskip-0.7mm\rangle}{\langle\hskip-0.7mm\langle\Delta\chi(t_{\mathrm{f}})\vert\Delta\chi(t_{\mathrm{f}})\rangle\hskip-0.7mm\rangle},\label{eq.19.2.2}
\end{gather} 
with $\Delta\mathpzc{F}=\mathpzc{F}\big(\chi^{(n)}(t_{\mathrm{f}})+\Delta\chi(t_{\mathrm{f}}),t_{\mathrm{f}}\big)-\mathpzc{F}\big(\chi^{(n)}(t_{\mathrm{f}}),t_{\mathrm{f}}\big)$. For a \textit{convex} final time-dependent objective as $\mathpzc{F}_{c}=- w_{0}\langle\hskip-0.7mm\langle\chi(t_{\mathrm{f}})\vert\Xi_{\,t_{\mathrm{f}}}\rangle\hskip-0.7mm\rangle$, we have $A=0$ and then $\bar{A}=0$ by setting $\zeta_{A}=0$. Since in this specific case $\sigma (t)$ is zero, then the Krotov method reduces to its first-order version---see Eq. \eqref{eq.18} with $\sigma(t)=0$.
  
By considering Eq. \eqref{eq.13} as the field-dependent function $\mathcal{G}_{f}$, Eq. \eqref{eq.18} leads to the following update equation: 
\begin{align}
\epsilon_{m}^{(n+1)}(t)=&\epsilon_{m}^{(\mathrm{ref})}(t)+\dfrac{f_{m}(t)}{ \hbar w_{m}}\Big\{\mathrm{Im}\langle\hskip-0.7mm\langle\Lambda(t)\, \vert\Big(\dfrac{\partial\mathbbmss{K}_{\bm{\epsilon}}}{\partial\epsilon_{m}}\Big)\Big| _{\bm{\epsilon}^{(n+1)}}\vert\chi^{(n+1)}(t)\rangle\hskip-0.7mm\rangle
\nonumber \\
&+\dfrac{\sigma(t)}{2}\mathrm{Im}\langle\hskip-0.7mm\langle\Delta\chi(t)\, \vert\Big(\dfrac{\partial\mathbbmss{K}_{\bm{\epsilon}}}{\partial\epsilon_{m}}\Big)\Big| _{\bm{\epsilon}^{(n+1)}}\vert\chi^{(n+1)}(t)\rangle\hskip-0.7mm\rangle\Big\}.
\label{eq.20}
\end{align} 
Following Ref. \cite{palao_quantum_2002,palao_optimal_2003}, throughout this paper we set $\epsilon_{m}^{(\mathrm{ref})}(t)=\epsilon_{m}^{(n)}(t)$. By this choice, the field-dependent functional \eqref{eq.13.1} vanishes when the fields approach their optimal values. Then, in this asymptotic limit the monotonic convergence of the total and the final time-dependent objectives, $\mathpzc{J}$ and $\mathpzc{F}$, are equivalent. The updated control field $\bm{\epsilon}^{(n+1)}(t)$ is considered as a guess field for the next iteration. Then, the above procedure is iterated until the algorithm achieves a desired convergence threshold.

A prohibitive issue to apply the second-order correction of the Krotov method, i.e., the second term in Eq. \eqref{eq.18} or the third term in Eq. \eqref{eq.20}, is to calculate the supremum over \textit{all} variations of the terminal dynamics $\{\Delta\chi(t_{\mathrm{f}})\}$ in order to obtain the constant coefficient $\bar{A}$ [Eqs. \eqref{eq.19.2.1} and \eqref{eq.19.2.2}]. A partial remedy for this is to replace $A$ in Eq. \eqref{eq.19.2.2} with a numerical ansatz as \cite{reich_monotonically_2012}
\begin{equation}
A^{(n+1)}=\dfrac{\Delta\mathpzc{F}^{(n+1)}+2\mathrm{Re}\langle\hskip-0.7mm\langle\Delta\chi^{(n+1)}(t_{\mathrm{f}})\vert\Lambda(t_{\mathrm{\mathrm{f}}})\rangle\hskip-0.7mm\rangle}{\langle\hskip-0.7mm\langle\Delta\chi^{(n+1)}(t_{\mathrm{f}})\vert\Delta\chi^{(n+1)}(t_{\mathrm{f}})\rangle\hskip-0.7mm\rangle},
\label{eq.21}
\end{equation} 
where $\Delta\mathpzc{F}^{(n+1)}=\mathpzc{F}\big(\chi^{(n+1)}(t_{\mathrm{f}}),t_{\mathrm{f}}\big)-\mathpzc{F}\big(\chi^{(n)}(t_{\mathrm{f}}),t_{\mathrm{f}}\big)$. However, the parameter $A^{(n+1)}$ depends on $\vert\Delta\chi^{(n+1)}(t_{\mathrm{f}})\rangle\hskip-0.7mm\rangle=\vert\chi^{(n+1)}(t_{\mathrm{f}})\rangle\hskip-0.7mm\rangle-\vert\chi^{(n)}(t_{\mathrm{f}})\rangle\hskip-0.7mm\rangle$ with an \textit{unknown} dynamics $\vert\chi^{(n+1)}(t_{\mathrm{f}})\rangle\hskip-0.7mm\rangle$ which is to be determined in the current iteration $n+1$. In order to resolve this difficulty, we can substitute the parameter $A^{(n)}$ calculated in the previous iteration into Eq. \eqref{eq.19.2.1}. Note that this procedure may compromise the monotonic convergence of the algorithm. In such case, this failed iteration needs to be repeated until monotonic convergence is guaranteed by considering $A^{(n+1)}$ instead of $A$ in Eq. \eqref{eq.19.2.1}. Another potential approach to resolve this issue is to set $\bar{A}=\zeta_{A}\geqslant 0$, and to find some $\zeta_{A}$ by trial and error such that the monotonic convergence can be retrieved \cite{reich_monotonically_2012}. 
  
\section{Application: Dynamical control of a Rydberg ion}
\label{sec:Application}

\begin{figure}
\includegraphics[scale=0.28]{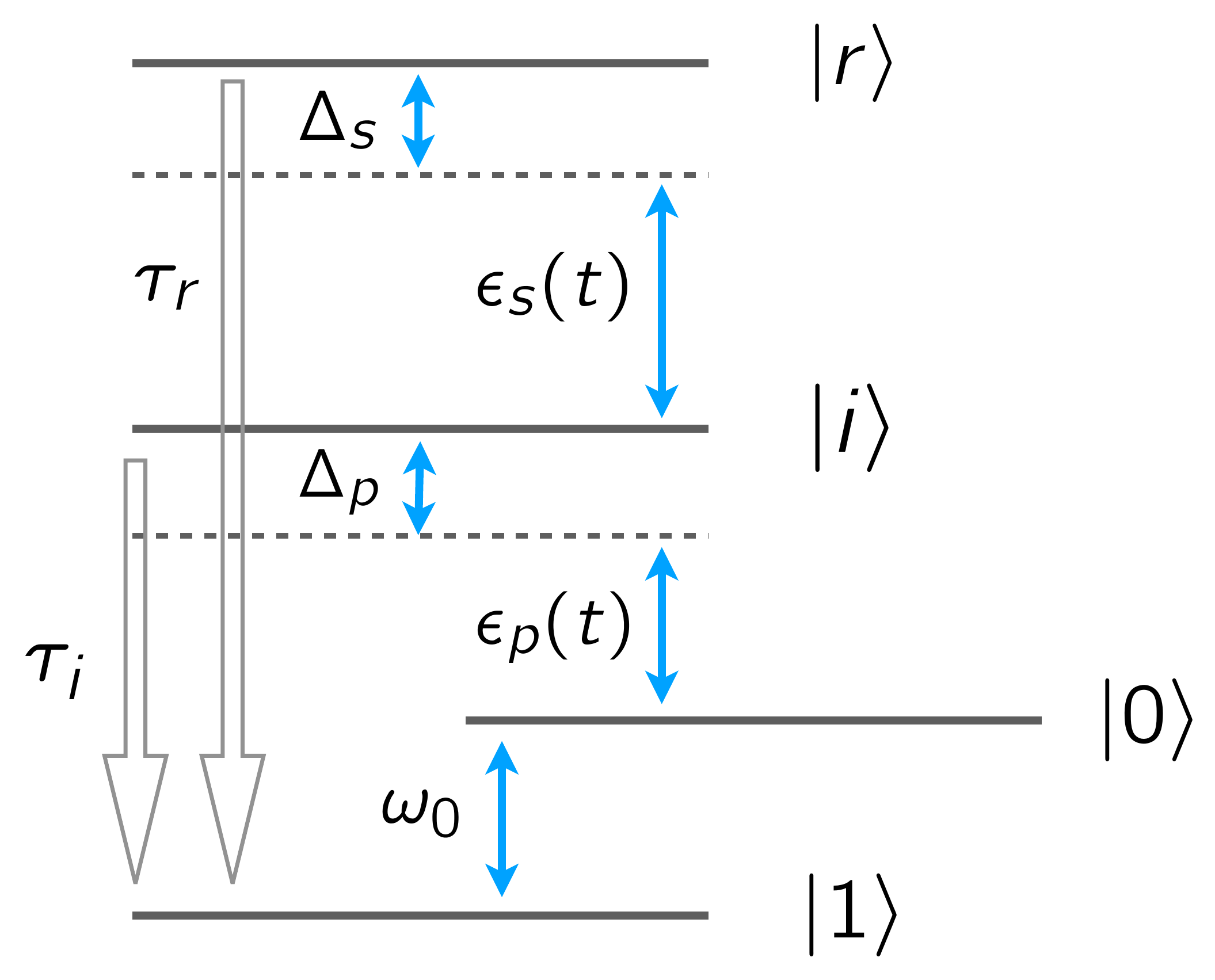}
\caption{Four energy levels of the Rydberg ion $^{88}\mathrm{Sr}^{+}$ driven by two laser pulses. Energy levels $\vert  0\rangle \equiv 4\mathrm{D}_{5/2}$ and $\vert  1\rangle \equiv 4\mathrm{S}_{1/2}$ as a register qubit with the transition frequency $\omega_{0}=(2\pi c)/\lambda_{0}$, in which $\lambda_{0}=674\,\mathrm{nm}$ ($c$ is the speed of light). The state $\vert  0\rangle$ is coupled to an intermediate state $\vert  i\rangle \equiv 6\mathrm{P}_{3/2}$ via a pump laser $\epsilon_{p}(t)$ with detuning $\Delta_{p}=40\pi\,\mathrm{MHz}$. The ionic transition $\vert i\rangle \leftrightarrow \vert r\rangle \equiv 42\mathrm{S}_{1/2}$ (Rydberg state) is driven by the Stokes laser $\epsilon_{s}(t)$ with detuning $\Delta_{s}=-\Delta_{p}$. The Rydberg state decays into the intermediate state within $\tau_{r}\approx 2.3\,\mu\mathrm{s}$. The decay time of the state $|i\rangle$ to the state $\vert 1\rangle$ is $\tau_{i}\approx35\,\mathrm{ns}$ \cite{higgins_coherent_2017}.}
\label{Fig-2:RydIon}
\end{figure}

Here we choose a related practical example to illustrate our dynamical control scheme. Rydberg neutral atoms and ions are appealing candidates for implementation of quantum computation with high fidelities \cite{Rydberg-RMP,madjarov_high-fidelity_2020, li_boost_2020, higgins_coherent_2017}. We consider a trapped Rydberg ion $^{88}\mathrm{Sr}^{+}$ \cite{higgins_coherent_2017} containing an energy level $\vert r\rangle \equiv 42\mathrm{S}_{1/2}$ as a Rydberg state. Only four energy levels of $^{88}\mathrm{Sr}^{+}$ have been shown in Fig. \ref{Fig-2:RydIon}. Two low-lying levels with the gap wavelength $\lambda_{0}=674\,\mathrm{nm}$ act as a register qubit, i.e., $\{\vert 0\rangle \equiv 4\mathrm{D}_{5/2},\,\vert 1\rangle \equiv 4\mathrm{S}_{1/2}\}$. The Rydberg state is addressable via two time-dependent lasers. The pump laser $\epsilon_{p}(t)$ couples the state $\vert 0\rangle$ to an intermediate state $\vert  i\rangle \equiv 6\mathrm{P}_{3/2}$. The trapped ion in the state $\vert i\rangle$ is also excited to the Rydberg state by a Stokes laser $\epsilon_{s}(t)$. We consider these control laser pulses as in the following:
\begin{equation}
\epsilon_{m}(t)=q_{m}(t)\, \cos (\omega_{m}t),\qquad m\in \{s,p\}, 
\end{equation}
where $q_{m}(t)$ and $\omega_{m}$ are the time-dependent amplitude and central frequency, respectively. Following Ref. \cite{goerz_optimizing_2015}, we consider an appropriate rotating frame defined by a suitable unitary operator defined as below in the logical basis $\{\vert 0\rangle,\,\vert 1\rangle ,\,\vert i\rangle ,\,\vert r\rangle\}$,
\begin{equation}
\mathpzc{U}(t)=\mathrm{diag}\big(1,1,e^{i\omega_{p}t},e^{i(\omega_{p}+\omega_{s})t}\big).
\label{eq.21-1}
\end{equation}
Hereon in this paper, we shall work in this particular rotating frame, without denoting a separate notation for quantities therein. After applying the rotating-wave approximation ($\omega_{s},\omega_{p}\gg \omega_{0}=(2\pi c)/\lambda_{0}$), the Hamiltonian of the system becomes \cite{higgins_coherent_2017, goerz_optimizing_2015}     
\begin{equation}
H_{S}(t)=\dfrac{\hbar}{2}
\begin{pmatrix}
0& 0& \Omega_{p}(t)& 0 \\
0& -2\omega_{0}& 0& 0 \\
\Omega_{p}(t)& 0& 2\Delta_{p}& -\Omega_{s}(t) \\
0& 0& -\Omega_{s}(t)& 2\big(\Delta_{p}+\Delta_{s} \big) \\
\end{pmatrix},
\label{eq.21-2}
\end{equation}
where the detuning $\Delta_{p}$ ($\Delta_{s}$) is defined as $\Delta_{p}=\omega_{i}-\omega_{p}$ ($\Delta_{s}=\omega_{r}-\omega_{i}-\omega_{s}$) and $(1/2)\Omega_{p}(t)=(1/2)\mu_{0i}\, q_{p}(t)$ ($(1/2)\Omega_{s}(t)=(1/2)\mu_{ir}\, q_{s}(t)$) is the time-dependent Rabi frequency of the pump (Stokes) laser pulse with the dipole moment $\mu_{0i}$ ($\mu_{ir}$) for the $\vert 0\rangle\leftrightarrow\vert i\rangle$  ($\vert i\rangle\leftrightarrow\vert r\rangle$) transition. The frequency of the former (latter) transition has been denoted by $\omega_{i}$ ($\omega_{s}$).  In the following we assume $\Delta_{p}=-\Delta_{s}=40\pi\,\mathrm{MHz}$ and refer to $\Omega_{p}(t)$ and $\Omega_{s}(t)$ (rather than $\epsilon_{p}(t)$ and $\epsilon_{s}(t)$) as the pump and Stokes laser pulses, respectively.

It has been known that the dominant effects of the surrounding environment of this trapped ion manifest as two spontaneous decays \citep{higgins_coherent_2017}: (a) $\vert i\rangle\rightarrow\vert 1\rangle$ with decay time $\tau_{i}\approx35\,\mathrm{ns}$ and (b) $\vert r\rangle\rightarrow\vert 1\rangle$ with decay time $\tau_{r}\approx 2.3\,\mu\mathrm{s}$. The associated Lindblad jump operators and rates of these processes are
\begin{gather}
L_{i}(t)=e^{-i\omega_{p}t}\vert 1\rangle\langle i\vert,\quad\gamma_{i}=1/\tau_{i},\label{eq.22}\\
L_{r}(t)=e^{-i(\omega_{p}+\omega_{s})t}\vert 1\rangle\langle r\vert,\quad\gamma_{r}=1/\tau_{r}.\label{eq.22-}
\end{gather}
From the dynamical equation \eqref{eq.6}, transformed into the rotating frame, with the time-dependent Hamiltonian \eqref{eq.21-2} and Lindblad operators \eqref{eq.22} and (\ref{eq.22-}) we can obtain a dynamical equation for the dynamics of this trapped ion. Note that the traceless operator basis $\{C_{\alpha}\}$ for this system is comprised of the generalized Gell-Mann basis matrices \cite{bertlmann_bloch_2008} for qudits. As an application of this dynamical equation we answer the following question in the next subsections by our coherent process control scheme: How can the dynamics of this Rydberg ion at a given final time be steered to a desired target process? 

In the following we consider three scenarios for the desired final process $\Xi_{\,t_{\mathrm{f}}}$. For the optimization we need the shape functions for the pump and Stokes lasers. Following Ref. \cite{goerz_robustness_2014}, here we choose the Blackman shape functions, given by
\begin{align}
f_{m}(t)=[1-g-\cos(k_{m}\pi t/t_{\mathrm{f}})+g\cos(l_{m}\pi t/t_{\mathrm{f}})]/2,
\label{eq.22-1}
\end{align}   
for $m\in\{s,p\}$, where $g=0.16$ and $k_{p}=4,\; l_{p}=8$ ($k_{s}=2,\; l_{s}=4$) for the pump (Stokes) laser. We also consider the following guess lasers to start the optimization process:
\begin{equation}
\Omega_{m}^{(0)}(t)=E_{m}f_{m}(t),\qquad m\in\{s,p\},
\label{eq.22-2}
\end{equation}   
where $E_{m}$ is the peak amplitude and we set $E_{m}=94\pi\,\mathrm{MHz}$ for both laser pulses \cite{higgins_coherent_2017}.

\ignore{
\begin{figure}[bp]
\includegraphics[scale=0.35]{OptimalPumpLaser-GateSimulation}
\caption{Optimal pump laser as a function of time for the phase gate simulation with $\varphi=\pi$ [see Eq. \eqref{eq.24}] and $t_{\mathrm{f}}=900 \,\mathrm{ns}$. Here $\lambda_{p}=0.01$. Inset indicates the spectrum of this laser. The laser and its spectrum have been shown in the rotating frame.}
\label{Fig-4}
\end{figure}
}

\begin{figure}[tp]
\includegraphics[width=\linewidth]{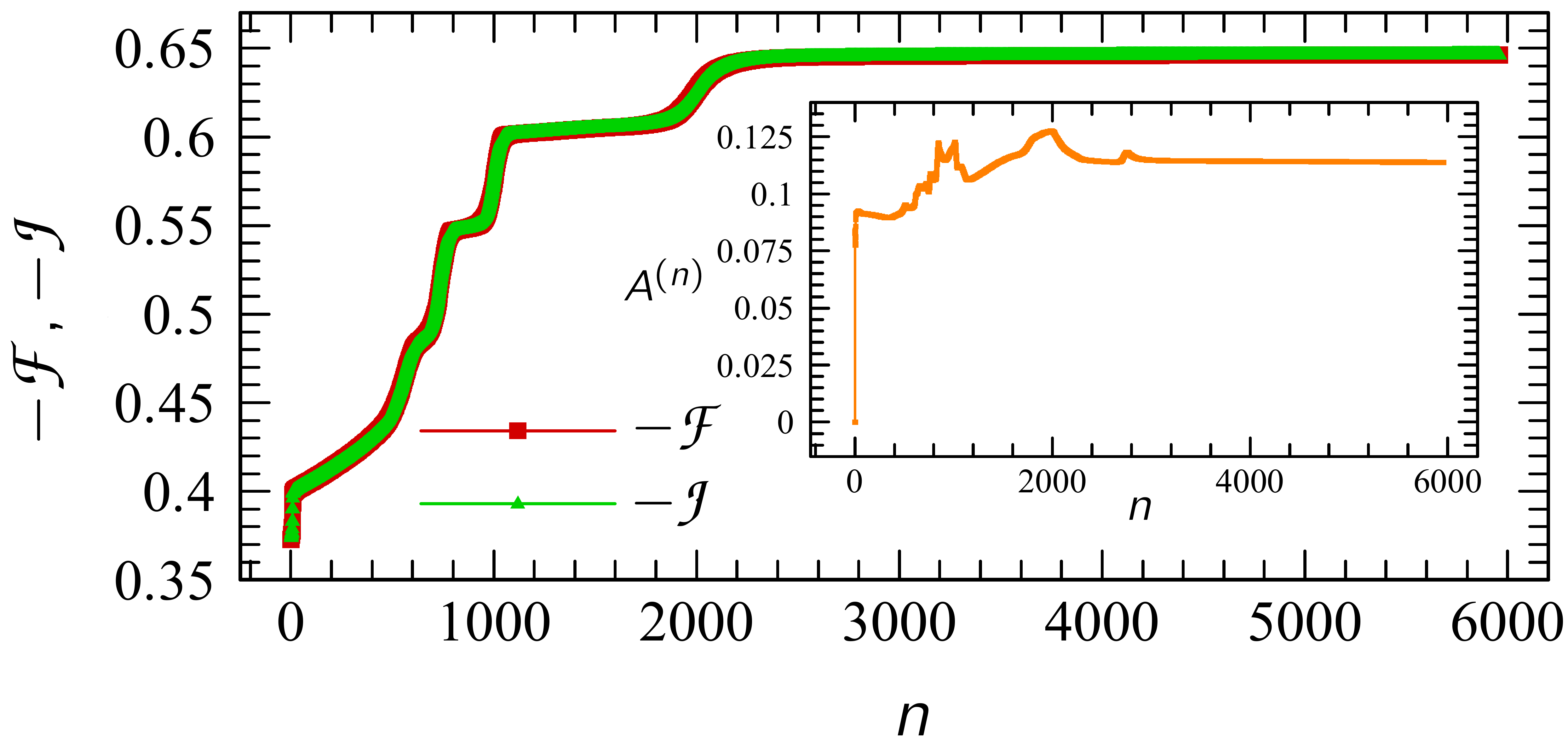}
\caption{Process fidelity $-\mathpzc{F}$ (red) and minus total functional $-\mathpzc{J}$ (green) vs. the iteration number $n$ for the phase gate simulation with $\varphi=\pi$ [see Eq. \eqref{eq.24}] and $t_{\mathrm{f}}=900 \,\mathrm{ns}$. Inset shows the control parameter $A^{(n)}$ defined in Eq. \eqref{eq.21} as a function of the iteration number $n$. Here $\zeta_{A}=0.01$ and $\zeta_{B}=0$ [see Eqs. (\ref{eq.19.1}) and (\ref{eq.19.2.1})].}
\label{Fig-3}
\end{figure}

\subsection{Scenario I: Gate simulation}
\label{subsec:GateSimu}

In the first scenario, the desired target process (in the rotating frame) is considered to be a specific unitary gate $O$, that is, 
\begin{equation}
[{\Xi}_{\,t_{\mathrm{f}}}^{(\mathrm{G})}]_{\alpha\beta}\equiv \mathrm{Tr}[OC_{\alpha}^{\dag}]\,\mathrm{Tr}[OC_{\beta}^{\dag}]^{*},
\label{eq.23}
\end{equation}
where $\alpha ,\beta \in\{1,\ldots, 16\}$. For example, here we aim to obtain the optimal Stokes and pump lasers which steer the dynamics of the register qubit ($\{\vert 0\rangle ,\vert 1\rangle\}$) to be similar to the phase gate applied at a specified final time (here $t_{\mathrm{f}}=900\;\mathrm{ns}$),  
\begin{equation}
O=\vert 0\rangle\langle 0\vert+e^{i\varphi}\vert 1\rangle\langle 1\vert+\vert i\rangle\langle i\vert+\vert r\rangle\langle r\vert. 
\label{eq.24}
\end{equation}
This operation acts trivially on the passive subspace $\{\vert i\rangle ,\vert r\rangle\}$ \cite{palao_quantum_2002}. Interestingly, Ref. \citep{higgins_coherent_2017} reports experimental realization of this gate on the register subspace by preparing the system as a pure state in this subspace. In contrast, our optimization scheme allows us to obtain any desired gate regardless of the preparation of the system in the register subspace.  

Figure \ref{Fig-3} shows the process fidelity $-\mathpzc{F}$ and the total functional $-\mathpzc{J}$ vs. the iteration number $n$ of the Krotov algorithm, both of which demonstrate similar behaviors. This figure indicates that the fidelity can improve only by a factor of $\approx\%27$ (up to iteration $5900$) through the optimization with the guess lasers chosen as in Eq. \eqref{eq.22-2}. Two plateaus are observed in the ranges $797<n<927$ and $1057<n<1893$, before the saturation of the fidelity at the value $\approx 0.646$. However, these plateaus are preliminary and the algorithm can still yield larger fidelities when it is given sufficiently larger iterations. This figure confirms that the iterative algorithm converges to the final objective $-\mathpzc{J}$ monotonically by the optimization procedure developed in Sec. \ref{sec:KrotovMethod}. This monotonic convergence occurs by incurring an extra numerical cost in updating the control laser pulses at each iteration, which is due to the last term in Eq. (\ref{eq.20})---see the inset of Fig. \ref{Fig-3}. 

\begin{figure}[tp]
\includegraphics[scale=0.33]{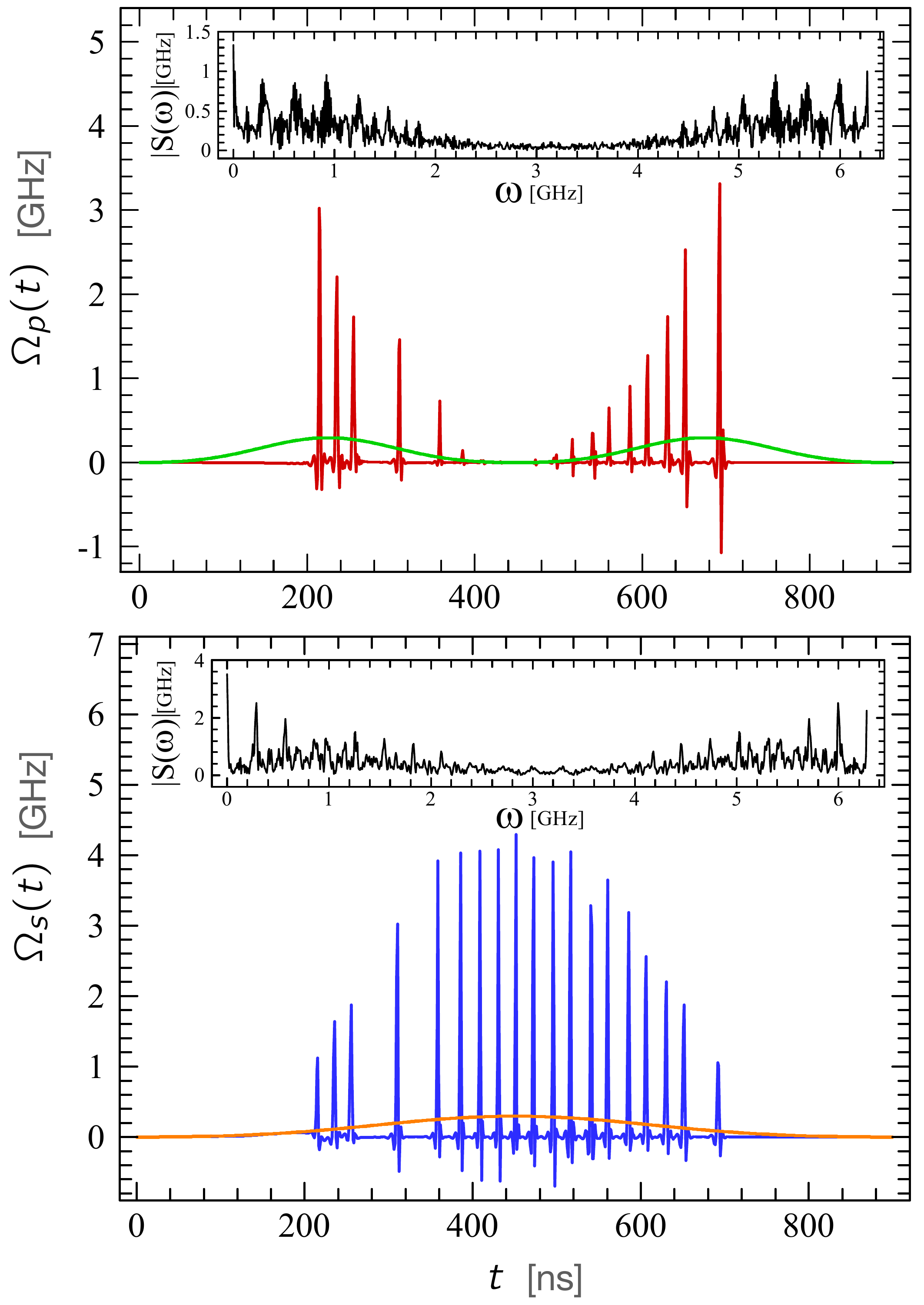}
\caption{Optimal pump laser (\textcolor{blue}{top}/red) and Stokes laser (\textcolor{blue}{bottom}/blue) vs. time for the phase-gate simulation with $\varphi=\pi$ [see Eq. \eqref{eq.24}] and $t_{\mathrm{f}}=900 \,\mathrm{ns}$. Insets indicate the spectra of these lasers. The lasers and their spectra have been shown in the rotating frame. The green and orange plots show the initial guess fields [see Eq. (\ref{eq.22-2})]. Here $w_{p}=w_{s}= 0.01$ [see Eq. (\ref{eq.13})].}
\label{Fig-5}
\end{figure}

The optimal pump and Stokes lasers have been shown in Fig. \ref{Fig-5}, respectively. It interesting to note that despite strike differences of the optimal and guess pulses, there is still a rough similarity between them---such a behavior has also been reported earlier in Ref. \cite{goerz_robustness_2014}. Note that in associated spectra the maximum peaks belong to the zero frequency, which in the lab frame correspond to the frequencies of the lasers whose detunings with the frequency of the $\vert 0\rangle\leftrightarrow\vert i\rangle$ and $\vert i\rangle\leftrightarrow\vert r\rangle$ transitions are given by $\Delta_{p}$  and $\Delta_{s}$, respectively.  

\subsection{Scenario II: Decoherence suppression}
\label{subsec:DecoSupp}

In this scenario, we are interested to see whether the pump and Stokes laser pulses applied to the trapped Rydberg ion can suppress the environment at a predetermined time $t_{\mathrm{f}}$ such that we have $\varrho_{S}(t_{\mathrm{f}}) = U_{S}^{\dag}(t_{\mathrm{f}})\varrho_{S}(0) U_{S}(t_{\mathrm{f}})$, where $U_{S}(t_{\mathrm{f}})$ is generated by the bare system Hamiltonian in the rotating frame, 
\begin{equation}
H_{S}=\hbar\big(-\omega_{0}\vert 1\rangle\langle 1\vert +\Delta_{p}\vert i\rangle\langle i\vert+(\Delta_{s}+\Delta_{p})\vert r\rangle\langle r\vert\big).
\label{eq.25}
\end{equation}
That is, the target process is given by 
\begin{align}
[{\Xi}^{(\mathrm{D})}_{\,t_{\mathrm{f}}}]_{\alpha\beta}\equiv \mathrm{Tr}[U_{S}(t_{\mathrm{f}}) \,C_{\alpha}^{\dag}]\,\mathrm{Tr}[U_{S}(t_{\mathrm{f}}) \, C_{\beta}^{\dag}]^{*},
\label{eq.26}
\end{align}
for $\alpha ,\beta \in\{1,\ldots, 16\}$. We set $t_{\mathrm{f}}=500\;\mathrm{ns}$ for the target time of this scenario. The environment acts on the ion through two quantum channels with decay times $\tau_{i}\approx 35\,\mathrm{ns}$ and $\tau_{r}\approx 2.3\,\mu\mathrm{s}$ (see Eq. \eqref{eq.22} for the jump rates and operators of these channels). Since $\tau_{i}<t_{\mathrm{f}}<\tau_{r}$, practically in this example we aim to suppress the detrimental effects of the spontaneous emission process $\vert i\rangle\rightarrow\vert 1\rangle$ at $t_{\mathrm{f}}=500\,\mathrm{ns}$.

\begin{figure}[tp]
\includegraphics[width=\linewidth]{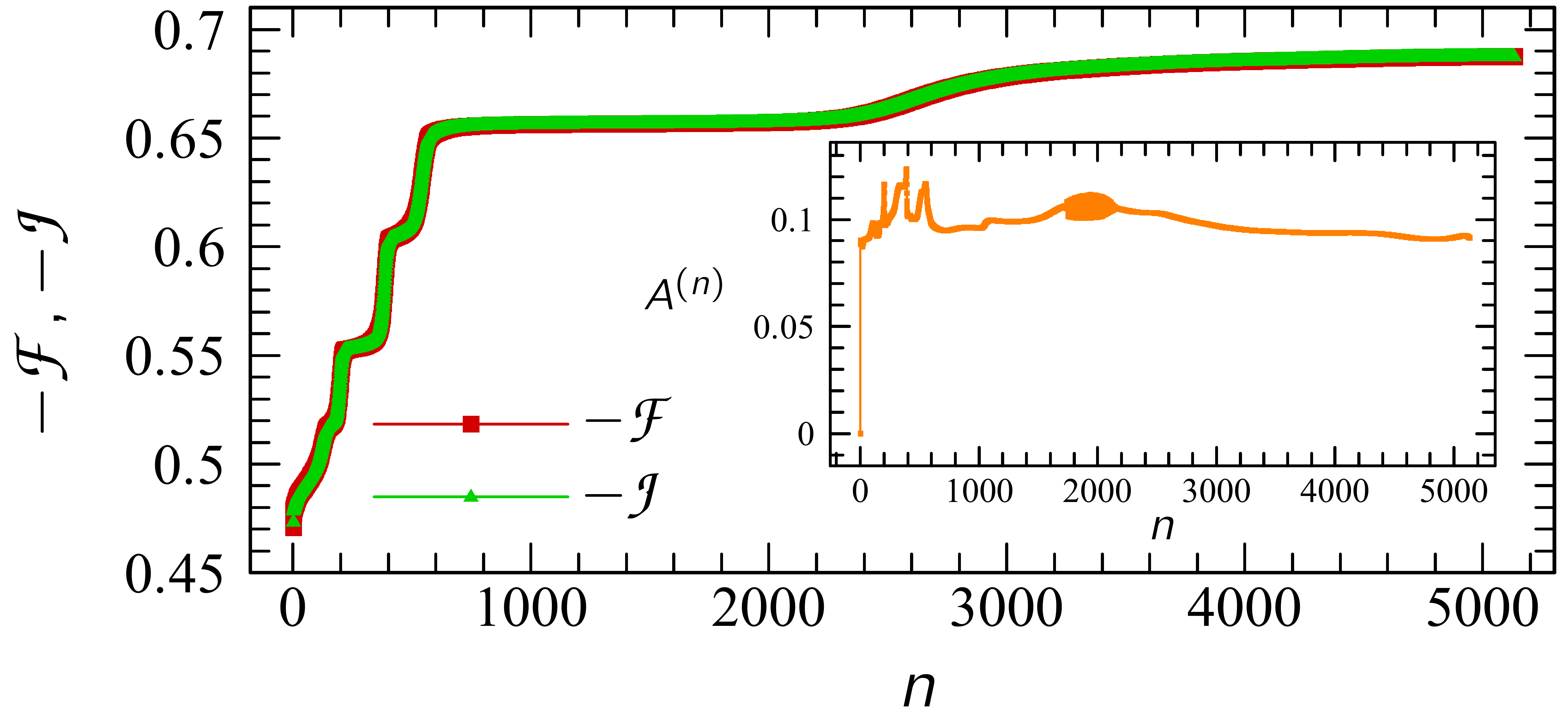}
\caption{Process fidelity $-\mathpzc{F}$ (red) and minus total functional $-\mathpzc{J}$ (green) vs. the iteration number $n$ for the decoherence suppression at $t_{\mathrm{f}}=500\,\mathrm{ns}$. Inset indicates the control parameter $A^{(n)}$ defined in Eq. \eqref{eq.21} as a function of the iteration number $n$. Here $\zeta_{A}=\zeta_{B}=0$ [see Eqs. (\ref{eq.19.1}) and (\ref{eq.19.2.1})].}
\label{Fig-6}
\end{figure}

The fidelity $-\mathpzc{F}$ as a function of the iteration number $n$ of the Krotov algorithm has been shown in Fig. \ref{Fig-6}. The fidelity reaches the value $\approx 0.687$ after $5133$ iterations. To go beyond this iteration, we have observed that the algorithm needs an exhaustive search on the space of the control parameter $A^{(n)}$ due to the ad hoc workaround for the numerical issue introduced in Sec. \ref{sec:KrotovMethod}. In the algorithm we have set $\zeta_{A}=0$. From Eqs. \eqref{eq.19.1}, \eqref{eq.19.2.1}, and \eqref{eq.20} this can lead to the Krotov method which is first order in $|\chi (t)\rangle\hskip-0.7mm\rangle$. We have also shown the control parameter $A^{(n)}$ as a function of the iteration number in the inset of Fig. \ref{Fig-6}. Except for the first iteration, in order to start the optimization process, the algorithm updates this parameter with nonzero values. The behavior seen at this inset (implying that $A^{(n)}>0$) in turn necessitates the second-order contribution in the updating equation \eqref{eq.20} to ensure monotonic convergence of the algorithm. A distinctive feature of this plot and also Fig. \ref{Fig-3} is the appearance of plateaus in some ranges of the iteration number, which indicate trappings in the algorithm (perhaps of the similar nature for trappings observed in Ref. \cite{ohtsuki_generalized_2004}).

\begin{figure}[tp]
\includegraphics[scale=0.33]{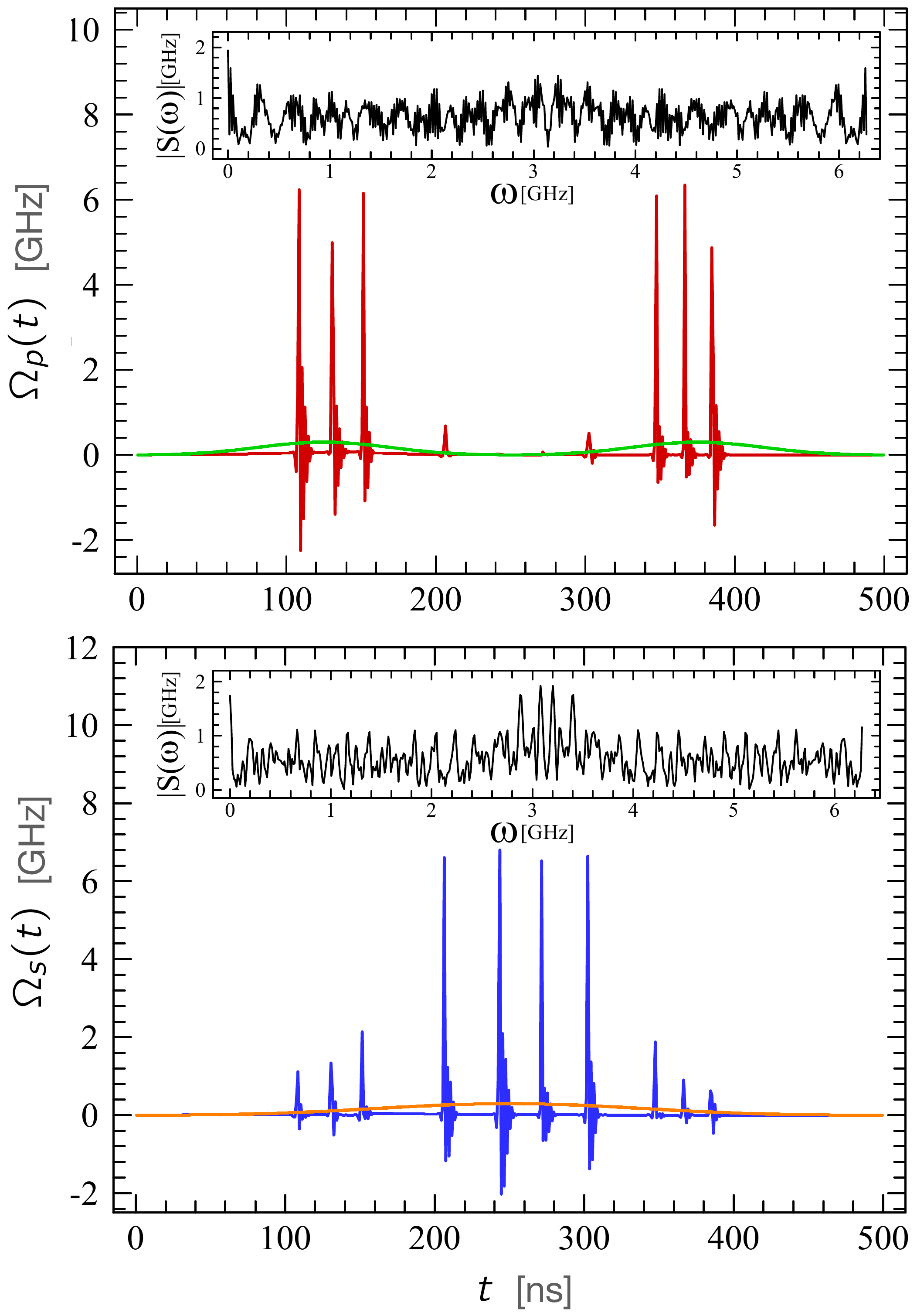}
\caption{Optimal pump laser (top/red) and Stokes laser (bottom/blue) vs. time for the decoherence suppression process [see. Eq. (\ref{eq.26})] at $t_{\mathrm{f}}=500 \,\mathrm{ns}$. Insets indicate the spectra of these lasers. The lasers and their spectra have been shown in the rotating frame. The green and orange plots show the initial guess fields [see Eq. (\ref{eq.22-2})]. Here $w_{p}=w_{s}= 0.01$ [see Eq. (\ref{eq.13})].}
\label{Fig-8}
\end{figure}

Figure \ref{Fig-8} indicates the optimal pump and Stokes laser pulses, respectively. Note the similarity and differences of the optimal and guess fields. The spectra of these fields have been depicted in the insets of these figures. We observe that a wide range of frequencies have significant contributions in the spectra of these optimal lasers. Since the bosonic environment contains a large number of frequencies, one may intuitively argue that a broadband spectra for the control laser fields may be needed to optimally alleviate the effect of the environment.

\subsection{Scenario III: Passive control of the environment}
\label{subsec:PassContEnv}

In this scenario we want to optimally control the environment by applying appropriate external fields on the system. In particular, we aim to modify the effect of the environment passively such that it looks differently (as we wish) to the system. This is particularly interesting noting that designing systems which can ``imposter'' another system has recently attracted much attention \cite{Campos_2017, McCaul_2020}. 

\begin{figure}[bp]
\includegraphics[width=\linewidth]{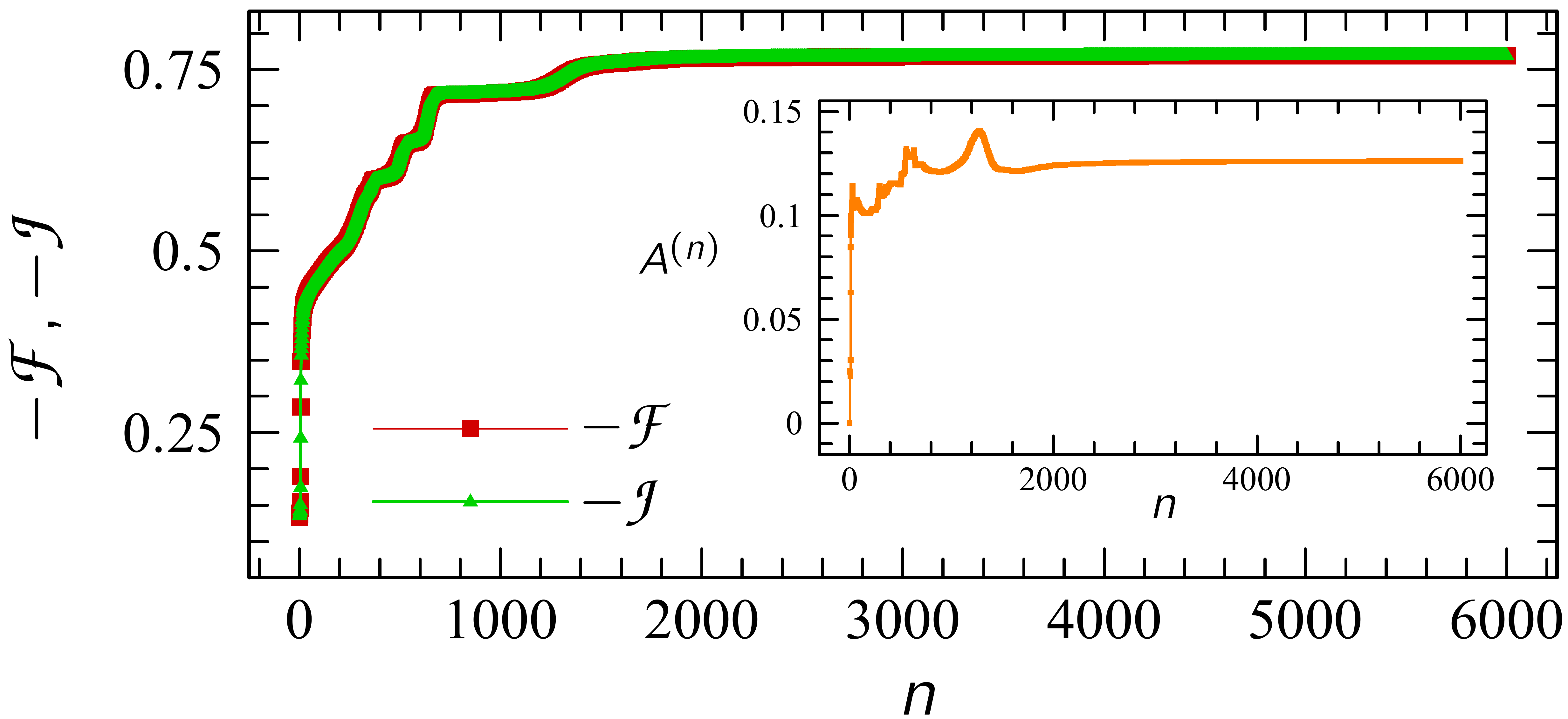}
\caption{Process fidelity (red) and minus the total functional (green) vs. iteration number $n$ for the passive control of the environment [see Eq. \eqref{eq.28}] and $t_{\mathrm{f}}=900 \,\mathrm{ns}$. Inset indicates the control parameter $A^{(n)}$ defined in Eq. \eqref{eq.21} as a function of the iteration number $n$. Here $\zeta_{A}=\zeta_{B}=0$ [see Eqs. (\ref{eq.19.1}) and (\ref{eq.19.2.1})].}
\label{Fig-9}
\end{figure}

In Fig. \ref{Fig-2:RydIon}, one of the effects of the environment on the Rydberg ion manifests as transferring population from $\vert i\rangle$ to $\vert 1\rangle$. Let us, for example, make this effect look like a depolarizing channel \cite{nielsen_quantum_2010} on the subspace $\{\vert 1\rangle ,\vert i\rangle\}$,
\begin{equation}
\mathpzc{D}^{(\mathrm{ch})}_{p}(\varrho)=[1-p(t)]\varrho + \big( p(t)/3\big)\textstyle{\sum_{\alpha=1}^{3}} \sigma_{\alpha} \varrho \sigma_{\alpha},
\label{eq.27}
\end{equation}
with $0\leqslant p(t)\, \leqslant 1$ and the operators $\sigma_{i}$ are defined as
\begin{align}
&\sigma_{1} \equiv\vert 1\rangle\langle i\vert+\vert i\rangle\langle 1\vert+\vert 0\rangle\langle 0\vert+\vert r\rangle\langle r\vert,
\nonumber\\
&\sigma_{2} \equiv -i\vert 1\rangle\langle i\vert+i\vert i\rangle\langle 1\vert+\vert 0\rangle\langle 0\vert+\vert r\rangle\langle r\vert,\\
&\sigma_{3} \equiv \vert 1\rangle\langle 1\vert -\vert i\rangle\langle i\vert+\vert 0\rangle\langle 0\vert+\vert r\rangle\langle r\vert.\nonumber
\end{align}
It can be seen that the reduction of these operators on the subspace $\{\vert 1\rangle ,\vert i\rangle\}$ acts similarly to the Pauli operators for a qubit. We assume here that $p(t)=(1-e^{-6t/\tau_{i}})/2$. This is indeed the error probability of a depolarizing channel acting on a qubit with the depolarizing time $\tau_{d}=\tau_{i}/6$ and the Kraus operators $W_{0}(\Delta t)=\sqrt{1-p(\Delta t)}\mathbbmss{I}_{2}-i\Delta tH_{0}$ and $W_{\alpha}(\Delta t)=\sqrt{p(\Delta t)/3}\sigma_{\alpha}$ ($H_{0}$ and $\Delta t$ are the free Hamiltonian of the qubit and a short time interval, respectively). The desired channel \eqref{eq.27} leads to the following target process at $t=t_{\mathrm{f}}$ in the generalized Gell-Mann basis for $N=4$, 
\begin{figure}[tp]
\includegraphics[scale=0.33]{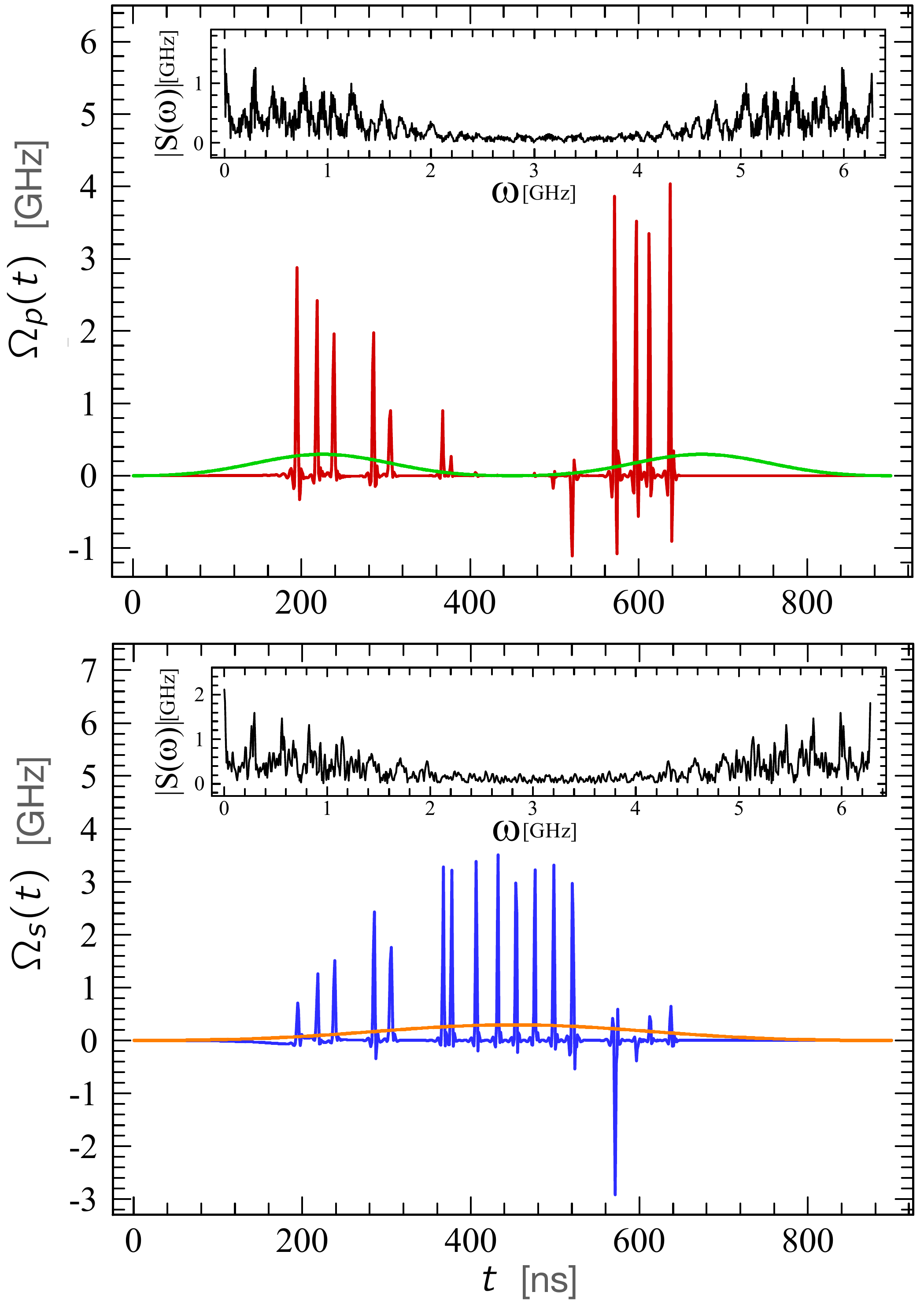}
\caption{Optimal pump laser (top/red) and Stokes laser (bottom/blue) vs. time for the passive control of the environment [see Eq. (\ref{eq.28})] at $t_{\mathrm{f}}=900 \,\mathrm{ns}$. Insets indicate the spectra of these lasers. The lasers and their spectra have been shown in the rotating frame. The green and orange plots show the initial guess fields [see Eq. (\ref{eq.22-2})]. Here $w_{p}=w_{s}= 0.01$ [see Eq. (\ref{eq.13})].}
\label{Fig-11}
\end{figure}
\begin{align}
[{\Xi}^{(\mathrm{ch})}_{\,t_{\mathrm{f}}}]_{\alpha\beta}=&\big\{\big(4-3p(t_{\mathrm{f}})\big)/2\big\}\delta_{\alpha16}\delta_{\beta16}+\big\{p(t_{\mathrm{f}})/6\big\}\big(2\delta_{\alpha 7}\delta_{\beta 7}
\nonumber\\
&+\delta_{\alpha 1}\delta_{\beta 1}+3\delta_{\alpha 6}\delta_{\beta 6}+\delta_{\alpha 11}\delta_{\beta 11}+2\delta_{\alpha 10}\delta_{\beta 10}
\nonumber\\
&+2\delta_{\alpha 1}\delta_{\beta 7}
+2\delta_{\alpha 1}\delta_{\beta 10}\big)
-\big\{\sqrt{6}p(t_{\mathrm{f}})/9\big\}\big(\delta_{\alpha 7}\delta_{\beta 11}
\nonumber\\
&+\delta_{\alpha 1}\delta_{\beta 11}
+\delta_{\alpha 10}\delta_{\beta 11}-3\delta_{\alpha 6}\delta_{\beta 16}\big)
+\big\{\sqrt{2}p(t_{\mathrm{f}})/3\big\}\nonumber\\
&\times\big(\delta_{\alpha 7}\delta_{\beta 16}
+\delta_{\alpha 1}\delta_{\beta 16}-\delta_{\alpha 6}\delta_{\beta 11}
+\delta_{\alpha 10}\delta_{\beta 16}\big)\nonumber\\
&+\big\{\sqrt{3}p(t_{\mathrm{f}})/9\big\}
\big(\delta_{\alpha 1}\delta_{\beta 6}+\delta_{\alpha 6}\delta_{\beta 7}
-3\delta_{\alpha 11}\delta_{\beta 16}
\nonumber\\
&+\delta_{\alpha 6}\delta_{\beta 10}\big)
+\alpha\leftrightarrow\beta,
\label{eq.28}
\end{align}
where $\alpha,\beta\in\{1,\ldots, 16\}$ and $\alpha\leftrightarrow\beta$ denotes terms similar to the previous after exchanging $\alpha \leftrightarrow \beta$. We set $t_{\mathrm{f}}=900\,\mathrm{ns}$ for the target time of this simulation. Figure \ref{Fig-9} indicates the process fidelity $-\mathpzc{F}$ and the total functional $-\mathpzc{J}$ vs. the iteration number $n$ of the Krotov algorithm. The optimization process improves the channel fidelity from $\approx 0.133$ to $\approx 0.769$ after performing $6000$ iterations. During the first $200$ iterations, the fidelity increases by a factor of $\%37$. Monotonic convergence of the total objective $-\mathpzc{J}$ is evident from this figure. We observe that in Figs. \ref{Fig-3}, \ref{Fig-6}, and \ref{Fig-9}, the convergence becomes slow when the algorithm approaches an optimal solution. In principle, this issue can be circumvented by combining the Krotov method and the quasi-Newton method \cite{Eitan_Optimal_Control}. However, we have not implemented this approach in this paper. The optimal laser pulses and their spectra have been shown in Fig. \ref{Fig-11}.   

\section{Summary and conclusions}
\label{sec:conc}

We have obtained an equation of motion for the process matrix associated with the dynamics of an open quantum system under the weak-coupling and Markovian assumptions. This equation is in the Lindblad form and resembles the master equation for the state or density matrix of the open system. Next, by using this equation, we have developed an open-system optimal control scheme where by local coherent manipulation of an open quantum system---through applying a control field---one can optimally implement quantum operations on the system under these conditions. The suitable optimal control field is given by minimizing a proper figure-of-merit, in which physical constraints have been included. This scheme can be straightforwardly extended to situations where applying a control field on the system may also affect how the environment acts on the system, and hence this scheme can enable various environment engineering scenarios. 

We have illustrated the utility of our scheme in three quantum control scenarios; decoherence suppression, gate simulation, and passive environment engineering. In the gate simulation scenario, the goal has been to force the system to evolve at a given time as closely as possible to a given unitary gate. In the decoherence suppression scenario, the objective has been to suppress as much as possible the effect of the interaction with the environment such that in a given time the evolution of the open system is simply given by its own Hamiltonian. The passive environment control scenario is an extension of the previous scenarios, in which simply by applying coherent control fields we have aimed to make the original environment look like another environment. Since these control scenarios are limited to coherent control of the system, i.e., without assuming the ability to manipulate the environment, they are subjective to the shape of applicable control fields, and may not achieve some operations with any desired high fidelity. However, our framework on its own is applicable to more general cases and can provide a feasible approach accessible with any given set of control operations.

\begin{acknowledgments}
V.R. acknowledges helpful discussions with Z. Nafari Qaleh and H. Yarlo. Partially supported by Sharif University of Technology's Office of Vice President for Research and Technology through Contract No. QA960512.
\end{acknowledgments}

\begin{widetext}
\onecolumngrid
\appendix
\section{Derivation of the process dynamical equation}
\label{app:a}

Here we present the derivation of the dynamical equation of $\chi (t)$ based on a formal approach. By using Eq. (\ref{eq.3}) and by differentiating the process matrix, we obtain the following linear differential equation:
\begin{equation}
 \dfrac{d\chi_{\alpha\beta}(t)}{dt}=\lim_{x \rightarrow 0}\dfrac{1}{x}\big(\chi_{\alpha\beta}(t+x)-\chi_{\alpha\beta}(t) \big) \equiv -\dfrac{i}{\hslash}\textstyle{\sum}_{\mu,\eta=1}^{N^{2}} [\mathpzc{K}]_{\alpha\beta,\mu\eta} \, \chi_{\mu\eta}(t),
\qquad \alpha,\beta\in\{1,\ldots,N^{2}\}.
\label{e.a1}
\end{equation}
From the definition of the operator $F_{\lambda}$, i.e., $[F_{\lambda}]_{\alpha\mu}\equiv\mathrm{Tr}[ C_{\alpha}^{\dag}C_{\lambda}C_{\mu}]$, we can write the time-dependent generator $\mathpzc{K}$ as
\begin{align} 
-\dfrac{i}{\hslash}[\mathpzc{K}]_{\alpha\beta,\mu\eta}=&\lim_{x\rightarrow 0}\frac{1}{x}\Big(\textstyle{\sum}_{\lambda,\gamma=1}^{N^{2}}\chi_{\lambda\gamma}(t+x,t) [ F_{\lambda}]_{\alpha\mu}
 [ F_{\gamma}]^{*}_{\beta\eta}-\delta_{\alpha\mu} \delta_{\beta\eta}\Big)
\label{a.e2}\\ 
=&\lim_{x\rightarrow 0}\dfrac{1}{x}\Big(\dfrac{1}{N}\chi_{N^{2}N^{2}}(t+x,t)\delta_{\beta\eta}\delta_{\alpha\mu}+\dfrac{1}{\sqrt{N}}\textstyle{\sum}_{\lambda=1}^{N^{2}-1}\big(\chi_{\lambda N^{2}}(t+x,t)\delta_{\beta\eta} [F_{\lambda}]_{\alpha\mu}
  +\chi_{N^{2}\lambda}(t+x,t)\delta_{\alpha\mu} [ F_{\lambda}]^{*}_{\beta\eta}\big) \nonumber\\
  &+\textstyle{\sum}_{\lambda,\gamma=1}^{N^{2}-1}\chi_{\lambda \gamma}(t+x,t) [ F_{\lambda}]_{\alpha\mu} [ F_{\gamma}]^{*}_{\beta\eta}-\delta_{\alpha\mu} \delta_{\beta\eta}\Big). 
\label{a.e3}
\end{align}
Now we introduce the time-dependent coefficients $a_{\lambda\gamma}(t)$ as
 \begin{align}
 a_{N^{2}N^{2}}(t) \equiv&\lim_{x\rightarrow 0} [\chi_{N^{2}N^{2}}(t+x,t)-N]/x,\nonumber\\
 a_{N^{2}\lambda}(t) \equiv&\lim_{x\rightarrow 0} \chi_{N^{2}\lambda}(t+x,t)/x,\qquad \lambda\in\{1,\ldots,N^{2}-1\},\nonumber\\
a_{\lambda\gamma}(t) \equiv&\lim_{x\rightarrow 0} \chi_{\lambda\gamma}(t+x,t)/x,\qquad \lambda,\gamma\in\{1,\ldots,N^{2}-1\}, \label{a.e4}
 \end{align}
 which lead to the following compact form for the components of the generator: 
\begin{align}
-\dfrac{i}{\hslash}[\mathpzc{K}] _{\alpha\beta,\mu\eta}=& \dfrac{1}{N}a_{N^{2}N^{2}}(t)\, \delta_{\beta\eta}\delta_{\alpha\mu}+\dfrac{1}{\sqrt{N}}\textstyle{\sum}_{\lambda=1}^{N^{2}-1}(a_{\lambda N^{2}}(t)\, \delta_{\beta\eta} [F_{\lambda}]_{\alpha\mu}+a_{N^{2}\lambda}(t)\, \delta_{\alpha\mu}[ F_{\lambda}]^{*}_{\beta\eta}) +\textstyle{\sum}_{\lambda,\gamma=1}^{N^{2}-1}a_{\lambda \gamma}(t) [F_{\lambda}]_{\alpha\mu}[ F_{\gamma}]^{*}_{\beta\eta}. 
 \label{a.e5}
 \end{align}

We shall now obtain the coefficients $a_{\lambda\gamma}(t)$ in terms of the operator basis of the Liouville space, i.e., $\{ C_{\alpha}\}_{\alpha=1}^{N^{2}}$. The trace-preserving property of the evolution map $\mathpzc{E}_{(t,t_{0})}$ leads to $\sum_{\alpha,\beta=1}^{N^{2}}\chi_{\alpha\beta}(t,t_{0}) \, C_{\beta}^{\dag} C_{\alpha}=\mathbbmss{I}_{S}$. Substituting $t_{0}\rightarrow t$ and $t\rightarrow t+x$ into the latter equation and then separating the terms including $K_{N^{2}}=\mathbbmss{I}_{S}/\sqrt{N}$ from the others yield
 \begin{align}
 \dfrac{1}{N}a_{N^{2}N^{2}}(t)\, \mathbbmss{I}_{S}+\dfrac{1}{\sqrt{N}}\textstyle{\sum}_{\lambda=1}^{N^{2}-1}\big(a_{\lambda N^{2}}(t)\, C_{\lambda}+a_{N^{2}\lambda}(t)\, C_{\lambda}^{\dag}\big)+\textstyle{\sum}_{\xi,\nu=1}^{N^{2}-1}a_{\xi\nu}(t)\, C_{\nu}^{\dag} C_{\xi}=0, 
 \label{a.e6}
 \end{align}
 where we have used the definition of the coefficients $a_{\xi\nu}(t)$ [Eqs. (\ref{a.e4})]. Now, we introduce the Hermitian operator 
  \begin{align}
H_{S}(t)=\dfrac{1}{2i}\big(M^{\dag}(t)-M(t)\big), 
 \label{a.e7}
 \end{align}
 where $M(t)=(\hbar/\sqrt{N}) \textstyle{\sum}_{\lambda=1}^{N^{2}-1}a_{\lambda N^{2}}(t)\, C_{\lambda}$, by which we can rewrite Eq. (\ref{a.e6}) as
\begin{align}
\dfrac{1}{N}a_{N^{2}N^{2}}(t)\, \mathbbmss{I}_{S}+\dfrac{2}{\hbar}M^{\dag}(t)-\dfrac{2i}{\hbar}H_{S}(t)+\textstyle{\sum}_{\xi,\nu=1}^{N^{2}-1}a_{\xi\nu}(t) \, C_{\nu}^{\dag} C_{\xi}=0. 
\label{a.e8}
\end{align}
After multiplying both sides of Eq. (\ref{a.e8}) by $C_{\lambda}$ ($\lambda\in\{1,\ldots,N^{2}-1\}$) and taking partial trace over system $S$, we obtain
\begin{align}
 a_{N^{2}\lambda}(t)=\dfrac{i}{\hbar}\sqrt{N}\,\mathrm{Tr} [H_{S}(t)\,C_{\lambda}] -\dfrac{\sqrt{N}}{2}\textstyle{\sum}_{\xi,\nu=1}^{N^{2}-1}a_{\xi\nu}(t)[F_{\xi}]_{\nu\lambda}.
 \label{a.e9}
 \end{align}
Note that we have supposed here that the operator basis $\{C_{\alpha}\}$ are traceless except that $C_{N^{2}}=(1/\sqrt{N})\mathbbmss{I}_{S}$, i.e., $\mathrm{Tr}[C_{\alpha}] =0$ for $\alpha\in\{1,\ldots, N^{2}-1\}$. Moreover, by taking partial trace over the system from both sides of Eq. (\ref{a.e6}), we obtain 
 \begin{align}
 a_{N^{2} N^{2}}(t)=-\textstyle{\sum}_{\nu,\xi=1}^{N^{2}-1}a_{\nu\xi}(t)\, \delta_{\nu\xi}.
 \label{a.e10}
 \end{align}
 Thus, by substituting Eqs. (\ref{a.e9}) and (\ref{a.e10}) into Eq. (\ref{a.e5}), the components of the generator $\mathpzc{K}$ can be obtained as 
  \begin{align}
-\dfrac{i}{\hslash}[ \mathpzc{K}]_{\alpha\beta,\mu\eta}=& -\dfrac{1}{N}\textstyle{\sum}_{\nu,\xi=1}^{N^{2}-1}a_{\nu\xi}(t)\, \delta_{\nu\xi}\delta_{\beta\eta}\delta_{\alpha\mu}+\textstyle{\sum}_{\lambda,\gamma=1}^{N^{2}-1}a_{\lambda\gamma}(t)\, [F_{\lambda}]_{\alpha\mu} [F_{\gamma}]_{\beta\eta}^{*}
  -\dfrac{i}{\hbar}\textstyle{\sum}_{\lambda=1}^{N^{2}-1}\mathrm{Tr}[C_{\lambda}^{\dag}H_{S}(t)] [F_{\lambda}]_{\alpha\mu} \delta_{\beta\eta} \nonumber\\ 
& +\dfrac{i}{\hbar}\textstyle{\sum}_{\lambda=1}^{N^{2}-1}\mathrm{Tr}[C_{\lambda}H_{S}(t)] [F_{\lambda}]_{\beta\eta}^{*}\delta_{\alpha\mu}
-\dfrac{1}{2}\textstyle{\sum}_{\nu,\xi=1}^{N^{2}-1}\textstyle{\sum}_{\lambda=1}^{N^{2}-1}a_{\nu\xi}(t) [F_{\xi}]_{\nu\lambda}^{*} [F_{\lambda}]_{\alpha\mu}\delta_{\beta\eta}
\nonumber \\
&-\dfrac{1}{2}\textstyle{\sum}_{\nu,\xi=1}^{N^{2}-1}\textstyle{\sum}_{\lambda=1}^{N^{2}-1} a_{\xi\nu}(t) [F_{\xi}]_{\nu\lambda}[F_{\lambda}]_{\beta\eta}^{*} \delta_{\alpha\mu}.
 \label{a.e11}
 \end{align}
By using Eqs. \eqref{a.e4} and \eqref{eq.2-2}, we can prove $\textstyle{\sum}_{\xi,\nu=1}^{N^{2}-1}\upsilon_{\xi}^{*}a_{\xi\nu}(t)\, \upsilon_{\nu}=\lim_{x\rightarrow 0}(1/x)\sum_{\lambda ,\mu}r_{\mu}^{2}\big| \sum_{\xi =1}^{N^{2}-1}\upsilon_{\xi}^{*}\mathrm{Tr}\big[\langle b_{\lambda}\vert U(t+x,t)\vert b_{\mu}\rangle C_{\xi}^{\dag}\big]\big|^{2}\geqslant 0$ for any $(N^{2}-1)-$dimensional vector $\upsilon$ and for any time $t$. Hence $(N^{2}-1)-$dimensional matrix ${a}(t)=[a_{\xi\nu}(t)]$ is positive semidefinite. By substituting Eq. (\ref{a.e11}) into Eq. (\ref{e.a1}), a set of coupled differential equations are obtained as
 \begin{align}
\dfrac{d\chi_{\alpha\beta}(t)}{dt}= &-\dfrac{1}{N}\textstyle{\sum}_{\xi,\nu=1}^{N^{2}-1}\textstyle{\sum}_{\mu,\eta=1}^{N^{2}}a_{\xi\nu}(t)\, \delta_{\nu\xi}\delta_{\beta\eta}\delta_{\alpha\mu} \, \chi_{\mu\eta}(t) + \textstyle{\sum}_{\xi,\nu=1}^{N^{2}-1}\textstyle{\sum}_{\mu,\eta=1}^{N^{2}}a_{\xi\nu}(t) [F_{\xi}]_{\alpha\mu} [F_{\nu}]_{\beta\eta}^{*}\, \chi_{\mu\eta}(t)\, \nonumber\\
&-\dfrac{i}{\hbar}\textstyle{\sum}_{\mu,\eta=1}^{N^{2}}\textstyle{\sum}_{\lambda=1}^{N^{2}}\mathrm{Tr}\big[C_{\lambda}^{\dag}H_{S}(t)\big]
[F_{\lambda}]_{\alpha\mu}\delta_{\beta\eta} \, \chi_{\mu\eta}(t) -\dfrac{1}{2}\textstyle{\sum}_{\xi,\nu=1}^{N^{2}-1}\textstyle{\sum}_{\mu,\eta=1}^{N^{2}}\textstyle{\sum}_{\lambda=1}^{N^{2}-1}a_{\xi\nu}(t) [F_{\nu}]_{\xi\lambda}^{*}[F_{\lambda}]_{\alpha\mu}\delta_{\beta\eta}
\, \chi_{\mu\eta}(t)
\nonumber \\
&-\dfrac{1}{2}\textstyle{\sum}_{\xi,\nu=1}^{N^{2}-1}\textstyle{\sum}_{\mu,\eta=1}^{N^{2}}\textstyle{\sum}_{\lambda=1}^{N^{2}-1}
a_{\xi\nu}(t) [F_{\xi}]_{\nu\lambda} [F_{\lambda}]_{\beta\eta}^{*} \delta_{\alpha\mu} \, \chi_{\mu\eta}(t) +\dfrac{i}{\hbar}\textstyle{\sum}_{\mu,\eta=1}^{N^{2}}\textstyle{\sum}_{\lambda=1}^{N^{2}}\mathrm{Tr}[C_{\lambda}H_{S}(t)] [F_{\lambda}]_{\beta\eta}^{*}\delta_{\alpha\mu} \, \chi_{\mu\eta}(t),
 \label{a.e12}
 \end{align}
where the upper limits of the summation over $\lambda$ in the third and sixth terms have been changed to $N^{2}$ because $H_{S}(t)$ is a traceless operator [see Eq. (\ref{a.e7})]. By using the orthonormality of the operator basis $\{C_{\alpha}\}$, the first term in Eq. (\ref{a.e12}) is absorbed into the forth and fifth terms by changing the upper limits of their summations over $\lambda$ to $N^{2}$. Hence Eq. (\ref{a.e12}) can be recast as
 \begin{align}
\dfrac{d\chi_{\alpha\beta}(t)}{dt}= &\textstyle{\sum}_{\xi,\nu=1}^{N^{2}-1}\textstyle{\sum}_{\mu,\eta=1}^{N^{2}}a_{\xi\nu}(t)[F_{\xi}]_{\alpha\mu} [F_{\nu}]_{\beta\eta}^{*} \, \chi_{\mu\eta}(t)-\dfrac{i}{\hbar}\textstyle{\sum}_{\mu,\eta=1}^{N^{2}}\textstyle{\sum}_{\lambda=1}^{N^{2}}\mathrm{Tr} [C_{\lambda}^{\dag}H_{S}(t)] [F_{\lambda}]_{\alpha\mu}\delta_{\beta\eta} \, \chi_{\mu\eta}(t)\, \nonumber\\
&-\dfrac{1}{2}\textstyle{\sum}_{\xi,\nu=1}^{N^{2}-1}\textstyle{\sum}_{\mu,\eta=1}^{N^{2}}\textstyle{\sum}_{\lambda=1}^{N^{2}}a_{\xi\nu}(t)
[F_{\nu}]_{\xi\lambda}^{*} [F_{\lambda}]_{\alpha\mu}\delta_{\beta\eta}
\, \chi_{\mu\eta}(t) -\dfrac{1}{2}\textstyle{\sum}_{\xi,\nu=1}^{N^{2}-1}\textstyle{\sum}_{\mu,\eta=1}^{N^{2}}\textstyle{\sum}_{\lambda=1}^{N^{2}}
a_{\xi\nu}(t) [F_{\xi}]_{\nu\lambda} [F_{\lambda}]_{\beta\eta}^{*}
 \delta_{\alpha\mu} \, \chi_{\mu\eta}(t)
 \nonumber \\
& +\dfrac{i}{\hbar}\textstyle{\sum}_{\mu,\eta=1}^{N^{2}}\textstyle{\sum}_{\lambda=1}^{N^{2}}\mathrm{Tr}[C_{\lambda}H_{S}(t)] [F_{\lambda}]_{\beta\eta}^{*}\delta_{\alpha\mu} \,\chi_{\mu\eta}(t).
 \label{a.e13}
 \end{align}

Equation (\ref{a.e13}) can still be brought into a more compact form. By expanding the Hermitian operator $H_{S}(t)$ in terms of the operator basis $\{C_{\lambda}\}_{\lambda=1}^{N^{2}} $ as $H_{S}(t)=\textstyle{\sum}_{\lambda=1}^{N^{2}}\mathrm{Tr}[C_{\lambda}^{\dag}H_{S}(t)] C_{\lambda}$, it is straightforward to prove the following equality:
\begin{equation}
\textstyle{\sum}_{\lambda=1}^{N^{2}}\mathrm{Tr}[C_{\lambda}^{\dag}H_{S}(t)] [F_{\lambda}]_{\alpha\mu}=\mathrm{Tr}[ C_{\alpha}^{\dag}H_{S}(t)C_{\mu}].
\label{a.e14}
\end{equation}
After some algebra, we also obtain another useful relation,  
\begin{align}
\textstyle{\sum}_{\lambda=1}^{N^{2}} [F_{\nu}]_{\xi\lambda}^{*} [F_{\lambda}]_{\alpha\mu} &=\textstyle{\sum}_{\lambda=1}^{N^{2}}\mathrm{Tr}[C_{\lambda}^{\dag} C_{\nu}^{\dag} C_{\xi}]\,\mathrm{Tr}[ C_{\alpha}^{\dag}C_{\lambda}C_{\mu}] =\mathrm{Tr}[C_{\alpha}^{\dag} C_{\nu}^{\dag} C_{\xi}C_{\mu}] \nonumber \\
&=\textstyle{\sum}_{\lambda=1}^{N^{2}}\mathrm{Tr}[C_{\lambda}^{\dag} C_{\nu} C_{\alpha}]^{*}\, \mathrm{Tr}[C_{\lambda}^{\dag} C_{\xi}C_{\mu}]
=\textstyle{\sum}_{\lambda=1}^{N^{2}}[F_{\nu}]_{\lambda\alpha}^{*} [F_{\xi}]_{\lambda\mu}
\nonumber \\
&=[F_{\nu}^{\dag}F_{\xi}]_{\alpha\mu},
\label{a.e15}
\end{align}
where we have used $ C_{\nu}^{\dag} C_{\xi}=\textstyle{\sum}_{\lambda=1}^{N^{2}}\mathrm{Tr}[C_{\lambda}^{\dag} C_{\nu}^{\dag} C_{\xi}] C_{\lambda} $ and $ C_{\xi}C_{\mu}=\textstyle{\sum}_{\lambda=1}^{N^{2}}\mathrm{Tr}[C_{\lambda}^{\dag} C_{\xi}C_{\mu}] C_{\lambda} $. Thus, by using Eqs. (\ref{a.e14}) and (\ref{a.e15}), one can get another form for Eq. (\ref{a.e13}) as
\begin{align}
\dfrac{d\chi_{\alpha\beta}(t)}{dt}= &\dfrac{i}{\hbar}\textstyle{\sum}_{\eta=1}^{N^{2}}\chi_{\alpha\eta}(t)\, \mathrm{Tr}[C_{\eta}^{\dag}H_{S}(t) C_{\beta}]
-\dfrac{i}{\hbar}\textstyle{\sum}_{\mu=1}^{N^{2}}\mathrm{Tr}\big[ C_{\alpha}^{\dag}H_{S}(t)C_{\mu}\big] \chi_{\mu\beta}(t)
 +\textstyle{\sum}_{\xi,\nu=1}^{N^{2}-1}\textstyle{\sum}_{\mu,\eta=1}^{N^{2}}a_{\xi\nu}(t)[F_{\xi}]_{\alpha\mu} \, \chi_{\mu\eta}(t)[F_{\nu}^{\dag}]_{\eta\beta}\nonumber\\
&-\dfrac{1}{2}\textstyle{\sum}_{\xi,\nu=1}^{N^{2}-1}\textstyle{\sum}_{\mu=1}^{N^{2}}a_{\xi\nu}(t)
 [F_{\nu}^{\dag}F_{\xi}]_{\alpha\mu} \, \chi_{\mu\beta}(t)-\dfrac{1}{2}\textstyle{\sum}_{\xi,\nu=1}^{N^{2}-1}\textstyle{\sum}_{\eta=1}^{N^{2}}
a_{\xi\nu}(t)\, \chi_{\alpha\eta}(t)[F_{\nu}^{\dag}F_{\xi}]_{\eta\beta}.
 \label{a.e16}
 \end{align}
Since the coefficient matrix $a(t)=[a_{\xi\nu}(t)]$ is a positive semidefinite matrix, we can diagonalize it via a time-dependent unitary matrix $u(t)$. Then, we have $u(t)a(t)u^{\dag}(t)=\gamma (t)$, where $\gamma (t)=\mathrm{diag}\big(\gamma_{\alpha}(t) \big)$ ($\alpha\in\{1,\ldots, N^{2}-1\}$) is the Lindblad coefficient matrix. Here, we define the time-dependent Lindblad operators as $A_{\lambda}(t)=\sum_{\xi=1}^{N^{2}-1}u_{\lambda\xi}^{\ast}(t) C_{\xi}$. Having these Lindblad coefficient matrix and operators  as well as introducing the $N^{2}-$dimensional matrices $\mathsf{H}_{S}(t)$ and $\mathsf{L}_{\alpha}(t)$, as Eq. \eqref{eq.7}, eventually lead to the dynamical equation of the dynamics (\ref{eq.6}). 

\section{Krotov method}
\label{app:b}

In this appendix, we adapt and extend the Krotov method, as discussed in Ref. \cite{reich_monotonically_2012}, to the process control of an open quantum system. For this purpose we first consider the process matrix $\chi (t)$ as an $N^{4}-$component vector $\vert\chi(t)\rangle\hskip-0.7mm\rangle$  and the field-dependent generator $\mathpzc{K}_{\,\,\bm{\epsilon}}$ as an $N^{4}-$dimensional matrix $\mathbbmss{K}_{\bm{\epsilon}}$ in the extended Hilbert space. This is equipped with the scalar product $\langle\hskip-0.7mm\langle\chi_{1}\vert\chi_{2}\rangle\hskip-0.7mm\rangle =\mathrm{Tr}[\chi_{1}^{\dag}\chi_{2}]$ for any $\vert\chi_{1}\rangle\hskip-0.7mm\rangle$ and $\vert\chi_{2}\rangle\hskip-0.7mm\rangle$ belonging to this extended space. Here, we summarize the problem of controlling the terminal process of an open system. One of the main goals of optimal control theory is to find the optimal fields $\bm{\epsilon}(t)=\{\epsilon_{m}(t)\}$ that minimize the following total objective functional:
\begin{equation}
\mathpzc{J}=\mathpzc{F}\big(\chi(t_{\mathrm{f}}),t_{\mathrm{f}}\big)+\textstyle{\int_{0}^{t_{\mathrm{f}}}}dt\,\mathcal{G}_{f}\big(\bm{\epsilon}(t),t\big),
\label{b.eq.1}
\end{equation}
with a final time-dependent function $\mathpzc{F}$ and a field-dependent function $\mathcal{G}_{f}$. In Eq. \eqref{b.eq.1} the notation $\chi(t_{\mathrm{f}})$ is to emphasize the dependence of the function $\mathpzc{F}$ to $\{\vert\chi(t_{\mathrm{f}})\rangle\hskip-0.7mm\rangle ,\langle\hskip-0.7mm\langle\chi(t_{\mathrm{f}})\vert\}$. The process also follows a dynamical equation as
\begin{equation}
\dfrac{d\vert\chi(t)\rangle\hskip-0.7mm\rangle}{dt}=-\dfrac{i}{\hbar}\mathbbmss{K}_{\bm{\epsilon}}\vert\chi(t)\rangle\hskip-0.7mm\rangle,\qquad\qquad\vert\chi(0)\rangle\hskip-0.7mm\rangle_{\alpha} =N\delta_{\alpha N^{4}},\quad\alpha\in\{1,\ldots, N^{4}\}.
\label{b.eq.2}
\end{equation} 
We proceed to solve this optimization problem via the Krotov method. From hereon and for brevity we may use the shorthand $\chi$ instead of $\{\vert\chi\rangle\hskip-0.7mm\rangle , \langle\hskip-0.7mm\langle\chi\vert\}$ as the dynamical variable of all intermediate time-dependent functions and omit the time variable $(t)$ from $\bm{\epsilon}(t)$, $\vert\chi (t)\rangle\hskip-0.7mm\rangle$ and $\langle\hskip-0.7mm\langle\chi (t)|$. Introducing an \textit{arbitrary} process-dependent and scalar function $\Upsilon(\chi ,t)$, we can rewrite the total functional \eqref{b.eq.1} as
\begin{equation}
\mathpzc{J}_{\,\Upsilon}=\mathpzc{M}_{\,\Upsilon}\big(\chi(t_{\mathrm{f}}),t_{\mathrm{f}}\big)-\Upsilon\big(\chi(0),0\big)
-\textstyle{\int_{0}^{t_{\mathrm{f}}}} dt\,\mathpzc{R}_{\, \Upsilon}\big(\chi(t),\bm{\epsilon}(t),t\big).
\label{b.eq.3}
\end{equation} 
where the modified final time-dependent function $\mathpzc{M}_{\,\Upsilon}$ and the intermediate time-dependent function $\mathpzc{R}_{\,\Upsilon}$ are given by
\begin{gather}
\mathpzc{M}_{\,\Upsilon}\big(\chi(t_{\mathrm{f}}),t_{\mathrm{f}}\big)=\mathpzc{F} \big(\chi(t_{\mathrm{f}}),t_{\mathrm{f}}\big)
+ \Upsilon\big(\chi(t_{\mathrm{f}}),t_{\mathrm{f}}\big), \label{b.eq.4}\\
\mathpzc{R}_{\, \Upsilon}(\chi,\bm{\epsilon},t) = -\mathcal{G}_{f}(\bm{\epsilon},t) +\dfrac{\partial\Upsilon}{\partial t}-\dfrac{i}{\hbar}\dfrac{\partial\Upsilon}{\partial\vert\chi\rangle\hskip-0.7mm\rangle}\mathbbmss{K}_{\bm{\epsilon}}\vert\chi\rangle\hskip-0.7mm\rangle+\dfrac{i}{\hbar}\langle\hskip-0.7mm\langle\chi\vert\mathbbmss{K}_{\bm{\epsilon}}^{\dag}\dfrac{\partial\Upsilon}{\partial\langle\hskip-0.7mm\langle\chi\vert}.
\label{b.eq.5}
\end{gather}
In fact, the dynamical constraint \eqref{b.eq.2} has been incorporated in the total functional $\mathpzc{J}_{\,\Upsilon}$ using the function $\Upsilon$. As we see later, the freedom in the choice of this function enables one to design a \textit{monotonically convergent} algorithm. That is the algorithm approaches the minimum of the modified functional $\mathpzc{J}_{\,\Upsilon}$ after each iteration,  
\begin{equation}
\mathpzc{J}_{\,\Upsilon}^{({n+1})}\leqslant \mathpzc{J}_{\,\Upsilon}^{({n})},\qquad\forall n\geqslant 0,
\label{b.eq.5-1}
\end{equation}
where 
\begin{equation}
\mathpzc{J}_{\,\Upsilon}^{(n)}=\mathpzc{M}_{\,\Upsilon}\big(\chi^{(n)}(t_{\mathrm{f}}),t_{\mathrm{f}}\big)-\Upsilon\big(\chi(0),0\big)
-\textstyle{\int_{0}^{t_{\mathrm{f}}}}dt\,\mathpzc{R}_{\,\Upsilon}\big(\chi^{(n)}(t),\bm{\epsilon}^{(n)}(t),t\big).
\label{b.eq.5-2}
\end{equation}
Choosing the process-dependent function $\Upsilon$ suitably is the principal core of the Krotov method. In summary, the Krotov method contains two successive steps, which we explain below.

\subsection{First step}

Having the control fields $\bm{\epsilon}^{(n)}$ leads to the dynamics $d\vert\chi^{(n)}\rangle\hskip-0.7mm\rangle/dt=(-i/\hbar)\mathbbmss{K}_{\bm{\epsilon}^{(n)}}\vert\chi^{(n)}\rangle\hskip-0.7mm\rangle$, with $\mathbbmss{K}$ evaluated at $\bm{\epsilon}^{(n)}$. These control fields $\bm{\epsilon}^{(n)}$ can be the guess fields at the beginning (when $n=0$) of the optimization process or the updated fields in the previous iteration (when $n>0$). In this step we fix and determine the arbitrary scalar function $\Upsilon$ such that the total functional $\mathpzc{J}_{\,\Upsilon}$ is \textit{maximized} over the dynamics $\chi^{(n)}$. This can be formulated in terms of the following conditions:
\begin{align}
&\mathrm{i.}\quad\mathpzc{R}_{\,\Upsilon}\big(\chi^{(n)},\bm{\epsilon}^{(n)},t\big)=\min_{\chi}\mathpzc{R}_{\,\Upsilon}\big(\chi,\bm{\epsilon}^{(n)},t\big)\qquad\qquad\forall t\in[0,t_{\mathrm{f}}],
\label{b.eq.6}\\
&\mathrm{ii.}\quad\mathpzc{M}_{\,\Upsilon}\big(\chi^{(n)}(t_{\mathrm{f}}),t_{\mathrm{f}}\big)=\max_{\chi(t_{\mathrm{f}})}\mathpzc{M}_{\,\Upsilon}\big(\chi(t_{\mathrm{f}}),t_{\mathrm{f}}\big).
\label{b.eq.7}
\end{align}
Note that the $\chi$ on the RHSs of the above equations do not need to satisfy Eq. (\ref{b.eq.2}); in this stage they are arbitrary dynamics. According to Eqs. \eqref{b.eq.6} and \eqref{b.eq.7}, the solution $\{\bm{\epsilon}^{(n)},\chi^{(n)}\}$ is the worst possible solution for the \textit{minimization} problem. In order to characterize this function, we assume that the RHS of dynamical equation \eqref{b.eq.2} and its Jacobian are bounded. Let us assume that the objective $\mathpzc{F}$ and the function $\mathcal{G}_{f}$ are bounded and twice differentiable. By adapting results of Ref. \cite{reich_monotonically_2012}, one can show that under these assumptions the following relation for the real-valued function $\Upsilon$ is a solution to the extremization problem posed by Eqs. \eqref{b.eq.6} and \eqref{b.eq.7}:
\begin{equation}
\Upsilon(\chi,t)=\langle\hskip-0.7mm\langle\chi\vert\Lambda\rangle\hskip-0.7mm\rangle+\langle\hskip-0.7mm\langle\Lambda\vert\chi\rangle\hskip-0.7mm\rangle+\dfrac{1}{2}\sigma (t)\, \langle\hskip-0.7mm\langle\Delta\chi\vert\Delta\chi\rangle\hskip-0.7mm\rangle,
\label{b.eq.8}
\end{equation}
where 
\begin{equation}
\vert\Delta\chi(t)\rangle\hskip-0.7mm\rangle=\vert\chi(t)\rangle\hskip-0.7mm\rangle-\vert\chi^{(n)}(t)\rangle\hskip-0.7mm\rangle
\end{equation}
is the change in the dynamics. Here the coefficients $\vert\Lambda(t)\rangle\hskip-0.7mm\rangle$ is defined as $\vert\Lambda(t)\rangle\hskip-0.7mm\rangle=\big(\partial\Upsilon/\partial\langle\hskip-0.7mm\langle\chi\vert\big)\big|_{\chi^{(n)}}$, and extrapolating the arguments of Ref. \cite{reich_monotonically_2012} the coefficient $\sigma(t)$ can be obtained as 
\begin{equation}
\sigma(t)=\tilde{a}(e^{\tilde{c}(t_{\mathrm{f}}-t)}-1)+\tilde{b},\qquad\qquad\tilde{a},\tilde{b},-\tilde{c}<0.
\label{b.eq.8-1}
\end{equation}
As a result, the extremization problem introduced in Eqs. \eqref{b.eq.6} and \eqref{b.eq.7} reduces to finding suitable coefficients $\tilde{a}$, $\tilde{b}$ and $\tilde{c}$. Before analytical calculation of these coefficients, we evaluate the necessary conditions to satisfy Eqs. \eqref{b.eq.6} and \eqref{b.eq.7}. First, we need to obtain a closed form for the function $\mathpzc{R}_{\,\Upsilon}$ [Eq. \eqref{b.eq.5}] by inserting $\Upsilon$ from Eq. \eqref{b.eq.8}. Note that Eq. \eqref{b.eq.8} also yields
\begin{gather}
\dfrac{\partial\Upsilon}{\partial \vert\chi\rangle\hskip-0.7mm\rangle}=\langle\hskip-0.7mm\langle\Lambda\vert +\dfrac{1}{2}\sigma(t)\, \langle\hskip-0.7mm\langle\Delta\chi\vert ,
\label{b.eq.8-2}\\
\dfrac{\partial\Upsilon}{\partial \langle\hskip-0.7mm\langle\chi\vert}=\vert\Lambda\rangle\hskip-0.7mm\rangle +\dfrac{1}{2}\sigma(t)\, \vert\Delta\chi\rangle\hskip-0.7mm\rangle, 
\label{b.eq.8-3}\\
\dfrac{\partial\Upsilon}{\partial t}=\langle\hskip-0.7mm\langle\chi\vert\dot{\Lambda}\rangle\hskip-0.7mm\rangle+\langle\hskip-0.7mm\langle\dot{\Lambda}\vert\chi\rangle\hskip-0.7mm\rangle+\dfrac{1}{2}\dot{\sigma} (t)\, \langle\hskip-0.7mm\langle\Delta\chi\vert\Delta\chi\rangle\hskip-0.7mm\rangle+\dfrac{1}{2}{\sigma} (t)\, \langle\hskip-0.7mm\langle\dot{\chi}^{(n)}\vert\Delta\chi\rangle\hskip-0.7mm\rangle+\dfrac{1}{2}{\sigma} (t)\, \langle\hskip-0.7mm\langle\Delta\chi\vert\dot{\chi}^{(n)}\rangle\hskip-0.7mm\rangle,
\label{b.eq.8-4}
\end{gather}
where dot is the shorthand for time derivative $(d/dt)$. By using Eqs. \eqref{b.eq.5} and \eqref{b.eq.8-2} -- \eqref{b.eq.8-4} along with $\vert\dot{\chi}^{(n)}\rangle\hskip-0.7mm\rangle=(-i/\hbar)\mathbbmss{K}_{\bm{\epsilon}^{(n)}}\vert\chi^{(n)}\rangle\hskip-0.7mm\rangle$ we obtain the modified function $\mathpzc{R}_{\,\Upsilon}$ as 
\begin{align}
\mathpzc{R}_{\,\Upsilon}(\chi,\bm{\epsilon},t)=&-\mathcal{G}_{f}(\bm{\epsilon},t)+\langle\hskip-0.7mm\langle\chi\vert\dot{\Lambda}\rangle\hskip-0.7mm\rangle 
+\langle\hskip-0.7mm\langle\dot{\Lambda}\vert\chi\rangle\hskip-0.7mm\rangle+\dfrac{1}{2}\dot{\sigma} (t)\, \langle\hskip-0.7mm\langle\Delta\chi\vert\Delta\chi\rangle\hskip-0.7mm\rangle-
\dfrac{i}{2\hbar}{\sigma} (t)\, \langle\hskip-0.7mm\langle{\chi}^{(n)}\vert\mathbbmss{K}_{\bm{\epsilon}^{(n)}}^{\dag}\vert\Delta\chi\rangle\hskip-0.7mm\rangle
+\dfrac{i}{2\hbar}{\sigma} (t)\, \langle\hskip-0.7mm\langle\Delta\chi\vert\mathbbmss{K}_{\bm{\epsilon}^{(n)}}\vert{\chi}^{(n)}\rangle\hskip-0.7mm\rangle
\nonumber\\
&-\dfrac{i}{\hbar}\langle\hskip-0.7mm\langle\Lambda\vert \mathbbmss{K}_{\bm{\epsilon}}\vert\chi\rangle\hskip-0.7mm\rangle-\dfrac{i}{2\hbar}\sigma(t)\, \langle\hskip-0.7mm\langle\Delta\chi\vert\mathbbmss{K}_{\bm{\epsilon}}\vert\chi\rangle\hskip-0.7mm\rangle
+\dfrac{i}{\hbar}\langle\hskip-0.7mm\langle\chi\vert \mathbbmss{K}_{\bm{\epsilon}}^{\dag}\vert\Lambda\rangle\hskip-0.7mm\rangle +\dfrac{i}{2\hbar}\sigma(t)\, \langle\hskip-0.7mm\langle\chi\vert\mathbbmss{K}_{\bm{\epsilon}}^{\dag}\vert\Delta\chi\rangle\hskip-0.7mm\rangle .
\label{b.eq.8-5}
\end{align}
Thus, the necessary condition to satisfy Eq. \eqref{b.eq.6}, $\big(\partial\mathpzc{R}_{\,\Upsilon}/\partial\langle\hskip-0.7mm\langle\chi\vert\big)\big|_{(\chi^{(n)},\bm{\epsilon}^{(n)})}=0$, reduces to
\begin{equation}
\dfrac{d\vert\Lambda (t) \rangle\hskip-0.7mm\rangle}{dt}=-\dfrac{i}{\hbar}\mathbbmss{K}_{\bm{\epsilon}^{(n)}}^{\dag}\vert\Lambda (t)\rangle\hskip-0.7mm\rangle.
\label{b.eq.9}
\end{equation}  
The boundary condition of this dynamical equation is obtained from Eq. \eqref{b.eq.7}. In a similar vein, the necessary condition for satisfying this relation, i.e., $\big(\partial\mathpzc{M}_{\,\Upsilon}/\partial\langle\hskip-0.7mm\langle\chi\vert\big)\big|_{\chi^{(n)}(t_{\mathrm{f}})}=0$, can be read from Eqs. \eqref{b.eq.4} and \eqref{b.eq.8} as
\begin{equation}
\vert\Lambda(t_{\mathrm{f}})\rangle\hskip-0.7mm\rangle=-\Big(\dfrac{\partial\mathpzc{F}}{\partial\langle\hskip-0.7mm\langle\chi(t_{\mathrm{f}})\vert}\Big)\Big| _{\chi^{(n)}(t_{\mathrm{f}})}.
\label{b.eq.10}
\end{equation}  

\subsubsection{Analytical calculation of $\sigma(t)$}

To calculate the constant parameters of the function $\sigma(t)$ in Eq. \eqref{b.eq.8-1}, one can first work out the following inequality equivalent to Eq. \eqref{b.eq.7}:
\begin{equation}
\Delta\mathpzc{M}_{\,\Upsilon}=\mathpzc{M}_{\,\Upsilon}\big(\chi^{(n)}(t_{\mathrm{f}})+\Delta\chi(t_{\mathrm{f}}),t_{\mathrm{f}}\big)-\mathpzc{M}_{\,\Upsilon}\big(\chi^{(n)}(t_{\mathrm{f}}),t_{\mathrm{f}}\big)< 0,
\label{in-eq}
\end{equation}
for all variations of the terminal dynamics $\Delta\chi(t_{\mathrm{f}})=\chi(t_{\mathrm{f}})-\chi^{(n)}(t_{\mathrm{f}})$. Using Eqs. \eqref{b.eq.4} and \eqref{b.eq.8}, and assuming $\langle\hskip-0.7mm\langle\Delta\chi(t_{\mathrm{f}})\vert\Delta\chi(t_{\mathrm{f}})\rangle\hskip-0.7mm\rangle\neq 0$, the inequality (\ref{in-eq}) gives
\begin{equation}
\dfrac{\Delta\mathpzc{F}+2\mathrm{Re}\langle\hskip-0.7mm\langle\Delta\chi(t_{\mathrm{f}})\vert\Lambda(t_{\mathrm{f}})\rangle\hskip-0.7mm\rangle}{\langle\hskip-0.7mm\langle\Delta\chi(t_{\mathrm{f}})\vert\Delta\chi(t_{\mathrm{f}})\rangle\hskip-0.7mm\rangle}+\dfrac{1}{2}\sigma(t_{\mathrm{f}})< 0,\qquad\qquad\forall \Delta\chi(t_{\mathrm{f}}),
\label{b.eq.11}
\end{equation} 
where
\begin{equation}
\Delta\mathpzc{F}=\mathpzc{F}\big(\chi^{(n)}(t_{\mathrm{f}})+\Delta\chi(t_{\mathrm{f}}),t_{\mathrm{f}}\big)-\mathpzc{F}\big(\chi^{(n)}(t_{\mathrm{f}}),t_{\mathrm{f}}\big).
\end{equation}
Thus, from Eq. \eqref{b.eq.8-1}, the worst-case scenario to fulfill the strict condition \eqref{b.eq.11} is
\begin{align}
&\qquad\quad\qquad 2A+\tilde{b}<0,
\label{b.eq.12}\\
&A=\sup_{\Delta\chi(t_{\mathrm{f}})}\dfrac{\Delta\mathpzc{F}+2\mathrm{Re}\langle\hskip-0.7mm\langle\Delta\chi(t_{\mathrm{f}})\vert\Lambda(t_{\mathrm{f}})\rangle\hskip-0.7mm\rangle}{\langle\hskip-0.7mm\langle\Delta\chi(t_{\mathrm{f}})\vert\Delta\chi(t_{\mathrm{f}})\rangle\hskip-0.7mm\rangle}.
\label{b.eq.13}
\end{align}
Now let us consider an inequality equivalent to the relation \eqref{b.eq.6}, i.e., 
\begin{equation}
\Delta\mathpzc{R}_{\,\Upsilon}=\mathpzc{R}_{\,\Upsilon}\big(\chi^{(n)}+\Delta\chi,\bm{\epsilon}^{(n)},t\big)-\mathpzc{R}_{\,\Upsilon}\big(\chi^{(n)},\bm{\epsilon}^{(n)},t\big)\geqslant 0,
\label{b.eq.14.1}
\end{equation}
for all $\Delta\chi =\chi-\chi^{(n)}$ and $t\in [0,t_{\mathrm{f}}]$. Employing Eqs. \eqref{b.eq.8-5} and \eqref{b.eq.9} we get the following equation for $\Delta\mathpzc{R}_{\,\Upsilon}$:
\begin{equation}
\Delta\mathpzc{R}_{\,\Upsilon}=\langle\hskip-0.7mm\langle\Delta\chi\vert\Delta\chi\rangle\hskip-0.7mm\rangle\Big(\dfrac{1}{2}\dot{\sigma}(t)+\dfrac{\sigma(t)}{\hbar}\dfrac{\mathrm{Im}\langle\hskip-0.7mm\langle\Delta\chi\vert\mathbbmss{K}_{\bm{\epsilon}^{(n)}}\vert\Delta\chi\rangle\hskip-0.7mm\rangle}{\langle\hskip-0.7mm\langle\Delta\chi\vert\Delta\chi\rangle\hskip-0.7mm\rangle}\Big).
\label{b.eq.14.2}
\end{equation}
From this equation and the fact that $x/2\geqslant -| x| \,\forall x\in\mathbbmss{R}$, we obtain a lower bound for $\Delta\mathpzc{R}_{\,\Upsilon}$ as follows:
\begin{gather}
\Delta\mathpzc{R}_{\,\Upsilon}\geqslant\langle\hskip-0.7mm\langle\Delta\chi\vert\Delta\chi\rangle\hskip-0.7mm\rangle\Big(\dfrac{1}{2}\dot{\sigma}(t)-|\sigma(t)|B \Big),
\label{b.eq.15}\\
B=\dfrac{2}{\hbar}\sup_{\lbrace\Delta\chi\rbrace;\; t\in[0,t_{\mathrm{f}}]}\Big|\dfrac{\mathrm{Im}\langle\hskip-0.7mm\langle\Delta\chi\vert\mathbbmss{K}_{\bm{\epsilon}^{(n)}}\vert\Delta\chi\rangle\hskip-0.7mm\rangle}{\langle\hskip-0.7mm\langle\Delta\chi\vert\Delta\chi\rangle\hskip-0.7mm\rangle}\Big|.
\label{b.eq.14}
\end{gather}
Here we assume $\langle\hskip-0.7mm\langle\Delta\chi\vert\Delta\chi\rangle\hskip-0.7mm\rangle\neq 0$, or equivalently $\Delta \chi \neq 0$. If it is demanded that this lower bound be positive 
\begin{equation}
\qquad\qquad\dfrac{1}{2}\dot{\sigma}(t)-|\sigma(t)| B> 0,
\label{b.eq.14.1}
\end{equation}
then the strict minimum condition for $\mathpzc{R}_{\,\Upsilon}$ at $\chi ^{(n)}$, i.e., $\Delta\mathpzc{R}_{\,\Upsilon}>0$, will hold. By using the fact that $| \sigma(t)|\geqslant \sigma(t),\;\forall t\in [0,t_{\mathrm{f}}]$, the condition \eqref{b.eq.14.1} can be reformulated as
\begin{equation}
\qquad\qquad\dfrac{1}{2}\dot{\sigma}(t)-\sigma(t)B> 0.
\label{b.eq.15}
\end{equation}
From Eq. \eqref{b.eq.8-1} we obtain the inequalities $\dot{\sigma}(t)\, \geqslant -\tilde{a}\tilde{c}$ and $\sigma (t)\, \leqslant \tilde{b}$, which in turn yield 
\begin{equation}
(1/2)\dot{\sigma}(t)-B\sigma(t)\, \geqslant -(\tilde{a}\tilde{c}/2)-B\tilde{b}.
\end{equation}
Then, imposing the following condition is equivalent to satisfying the condition \eqref{b.eq.15}:
\begin{equation}
-\dfrac{1}{2}\tilde{a}\tilde{c}-B\tilde{b}> 0,
\label{b.eq.16}
\end{equation}
which does not depend on time $t$.

We proceed here to find the parameters $\tilde{a}$, $\tilde{b}$, and $\tilde{c}$ of the inequalities \eqref{b.eq.12} and \eqref{b.eq.16}. We set $\tilde{a}$ and $\tilde{b}$ as
\begin{equation}
\tilde{a}=\tilde{b}=-\bar{A},\qquad\bar{A}=\max \{\zeta_{A},2A+\zeta_{A}\},
\label{b.eq.17}
\end{equation}
where $\zeta_{A}>0$. If $\bar{A}=\zeta_{A}$ ($\bar{A}=2A+\zeta_{A}$), then we have $2A-\zeta_{A} <0$ ($-\zeta_{A}<0$) for the inequality \eqref{b.eq.12}. Therefore, the last inequality is satisfied with the choice of $\tilde{a}$ and $\tilde{b}$ as in Eq. \eqref{b.eq.17}. By this choice the inequality \eqref{b.eq.16} also becomes
\begin{equation}
\tilde{c}+2B>0.
\label{b.eq.17-1}
\end{equation}
Since $B\geqslant 0$ [Eq. \eqref{b.eq.14}], then one solution for the parameter $\tilde{c}$ is
\begin{equation}
\tilde{c}=\zeta_{B},\qquad\zeta_{B}>0.
\label{b.eq.17-2}
\end{equation} 
By substituting Eqs. \eqref{b.eq.17} and \eqref{b.eq.17-2} into Eq. \eqref{b.eq.8-1}, we obtain the following form for $\sigma (t)$:
\begin{equation}
\sigma (t)=-\bar{A}e^{\zeta_{B}(t_{\mathrm{f}}-t)}.
\label{b.eq.17-3}
\end{equation}
Note that we can also set the parameters $\zeta_{A}$ and $\zeta_{B}$ to zero. For more details on this particular choice, see the end of this section. 

\subsection{Second step}

The dynamics-dependent function $\Upsilon$ obtained in the previous step [Eq. \eqref{b.eq.8}] with the choice of $\sigma (t)$ as in Eq. \eqref{b.eq.17-3} enables us to find a local minimum for the total functional $\mathpzc{J}_{\,\Upsilon}$ [Eq. \eqref{b.eq.3}] with respect to the control fields $\bm{\epsilon}(t)$. According to Eq. \eqref{b.eq.3}, this implies that
\begin{equation}
\bm{\epsilon}^{(n+1)}=\mathrm{arg}\{\max_{\bm{\epsilon}}\mathpzc{R}_{\,\Upsilon}(\chi,\bm{\epsilon},t)\},\qquad\qquad\forall t\in[0,t_{\mathrm{f}}].
\label{b.eq.18}
\end{equation}  
The optimal field $\bm{\epsilon}^{(n+1)}$ obtained in this step needs to be compatible with the dynamics $\chi^{(n+1)}$ given by
\begin{equation}
\dfrac{d\vert\chi^{(n+1)}\rangle\hskip-0.7mm\rangle}{dt}=-\dfrac{i}{\hbar}\mathbbmss{K}_{\bm{\epsilon}^{(n+1)}}\vert\chi^{(n+1)}\rangle\hskip-0.7mm\rangle,\qquad\qquad\vert\chi^{(n+1)}(0)\rangle\hskip-0.7mm\rangle_{\alpha}=N\delta_{\alpha N^{4}},\quad\alpha\in\{1,\ldots, N^{4}\}.
\label{b.eq.19}
\end{equation}
Thus, the maximization problem \eqref{b.eq.18} and the dynamical equation \eqref{b.eq.19} should be solved simultaneously. We first consider one of the necessary conditions for the maximization problem \eqref{b.eq.18}, i.e., $\big(\partial\mathpzc{R}_{\,\Upsilon}/\partial\epsilon_{m}\big)\big|_{(\chi^{(n+1)},\bm{\epsilon}^{(n+1)})}=0$. Employing Eq. \eqref{b.eq.8-5} one can obtain an expression for the partial derivative $\partial\mathpzc{R}_{\,\Upsilon}/\partial\epsilon_{m}$ as
\begin{align}
\dfrac{\partial\mathpzc{R}_{\,\Upsilon}}{\partial\epsilon_{m}}=-\dfrac{\partial\mathcal{G}_{f}}{\partial\epsilon_{m}}
-\dfrac{i}{\hbar}\langle\hskip-0.7mm\langle\Lambda\vert\Big(\dfrac{\partial\mathbbmss{K}_{\bm{\epsilon}}}{\partial\epsilon_{m}}\Big)\vert\chi\rangle\hskip-0.7mm\rangle-\dfrac{i}{2\hbar}\sigma(t)\, \langle\hskip-0.7mm\langle\Delta\chi\vert\Big(\dfrac{\partial\mathbbmss{K}_{\bm{\epsilon}}}{\partial\epsilon_{m}}\Big)\vert\chi\rangle\hskip-0.7mm\rangle
+\dfrac{i}{\hbar}\langle\hskip-0.7mm\langle\chi\vert \Big(\dfrac{\partial\mathbbmss{K}_{\bm{\epsilon}}^{\dag}}{\partial\epsilon_{m}}\Big)\vert\Lambda\rangle\hskip-0.7mm\rangle +\dfrac{i}{2\hbar}\sigma(t)\, \langle\hskip-0.7mm\langle\chi\vert\Big(\dfrac{\partial\mathbbmss{K}_{\bm{\epsilon}}^{\dag}}{\partial\epsilon_{m}}\Big)\vert\Delta\chi\rangle\hskip-0.7mm\rangle. 
\label{b.eq.19-1}
\end{align}
Then, the local maximum condition for the function $\mathpzc{R}_{\,\Upsilon}$ at $\bm{\epsilon}^{(n+1)}$, i.e., $\big(\partial\mathpzc{R}_{\,\Upsilon}/\partial\epsilon_{m}\big)\big|_{(\chi^{(n+1)},\bm{\epsilon}^{(n+1)})}=0$, becomes
\begin{align}
\Big(\dfrac{\partial\mathcal{G}_{f}}{\partial\epsilon_{m}}\Big)\Big|_{\bm{\epsilon}^{(n+1)}}=\dfrac{2}{\hbar}\mathrm{Im}\Big\{\langle\hskip-0.7mm\langle\Lambda(t)\, \vert\Big(\dfrac{\partial\mathbbmss{K}_{\bm{\epsilon}}}{\partial\epsilon_{m}}\Big)\Big| _{\bm{\epsilon}^{(n+1)}}\vert\chi^{(n+1)}(t)\rangle\hskip-0.7mm\rangle\Big\}
+\dfrac{\sigma(t)}{\hbar}\mathrm{Im}\Big\{\langle\hskip-0.7mm\langle\Delta\chi^{(n+1)}(t)\, \vert\Big(\dfrac{\partial\mathbbmss{K}_{\bm{\epsilon}}}{\partial\epsilon_{m}}\Big)\Big| _{\bm{\epsilon}^{(n+1)}}\vert\chi^{(n+1)}(t)\rangle\hskip-0.7mm\rangle\Big\},
\label{b.eq.19-2}
\end{align}
where $\vert\Delta\chi^{(n+1)}(t)\rangle\hskip-0.7mm\rangle=\vert\chi^{(n+1)}(t)\rangle\hskip-0.7mm\rangle-\vert\chi^{(n)}(t)\rangle\hskip-0.7mm\rangle$. In this work, we restrict ourself to the field-dependent function $\mathcal{G}_{f}$ similar to Eq. \eqref{eq.13}, which are quadratic in the fields. We also assume that the field-dependent generator $\mathbbmss{K}_{\bm{\epsilon}}$ depends on the fields $\bm{\epsilon}$ linearly through the field-system interaction Hamiltonian $V_{\mathrm{field}}(t)$. Having such constraint and field-dependent generator as well as using Eq. \eqref{b.eq.19-1}, we obtain another necessary condition for the local maximum of the function $\mathpzc{R}_{\,\Upsilon}$ at $\bm{\epsilon}^{(n+1)}$, i.e., $\big(\partial^{2}\mathpzc{R}_{\,\Upsilon}/\partial\epsilon_{m}^{2}\big)\big|_{(\chi^{(n+1)},\bm{\epsilon}^{(n+1)})}< 0$, as follows:
\begin{equation}
 w_{m} / f_{m}(t)>0.
\label{b.eq.20}
\end{equation}

After these two steps, the change of the total functional $\mathpzc{J}_{\,\Upsilon}$ by iteration becomes
\begin{align}
\Delta\mathpzc{J}_{\,\Upsilon}^{(n+1)}=&\mathpzc{J}_{\,\Upsilon}^{(n+1)}-\mathpzc{J}_{\,\Upsilon}^{(n)}=\mathpzc{M}_{\,\Upsilon}\big(\chi^{(n+1)}(t_{\mathrm{f}}),t_{\mathrm{f}}\big)-\mathpzc{M}_{\,\Upsilon}\big(\chi^{(n)}(t_{\mathrm{f}}),t_{\mathrm{f}}\big)
\nonumber\\
&-\textstyle{\int_{0}^{t_{\mathrm{f}}}}dt\,\big\{\mathpzc{R}_{\,\Upsilon}\big(\chi^{(n+1)},\bm{\epsilon}^{(n+1)},t\big)-\mathpzc{R}_{\,\Upsilon}\big(\chi^{(n+1)},\bm{\epsilon}^{(n)},t\big)+\mathpzc{R}_{\,\Upsilon}\big(\chi^{(n+1)},\bm{\epsilon}^{(n)},t\big)-\mathpzc{R}_{\,\Upsilon}\big(\chi^{(n)},\bm{\epsilon}^{(n)},t\big)\big\},
\label{b.eq.21}
\end{align}
where for writing the RHS of the second equality we have used Eq. \eqref{b.eq.5-2}. Equations \eqref{b.eq.6} and \eqref{b.eq.7} lead to the inequalities $\mathpzc{R}_{\,\Upsilon}\big(\chi^{(n+1)},\bm{\epsilon}^{(n)},t\big)\geqslant \mathpzc{R}_{\,\Upsilon}\big(\chi^{(n)},\bm{\epsilon}^{(n)},t\big),\;\forall t\in [0,t_{\mathrm{f}}]$ and $\mathpzc{M}_{\,\Upsilon}\big(\chi^{(n+1)}(t_{\mathrm{f}}),t_{\mathrm{f}}\big)\leqslant \mathpzc{M}_{\,\Upsilon}\big(\chi^{(n)}(t_{\mathrm{f}}),t_{\mathrm{f}}\big)$. We also obtain the inequality $\mathpzc{R}_{\,\Upsilon}\big(\chi^{(n+1)},\bm{\epsilon}^{(n+1)},t\big)\geqslant \mathpzc{R}_{\,\Upsilon}\big(\chi^{(n+1)},\bm{\epsilon}^{(n)},t\big),\,\forall t\in [0,t_{\mathrm{f}}]$ from Eq. \eqref{b.eq.18}. From these inequalities and Eq. \eqref{b.eq.21}, the monotonic decrease of the modified total functional $\mathpzc{J}_{\,\Upsilon}$ with respect to the iteration number, i.e., $\Delta\mathpzc{J}_{\,\Upsilon}^{(n+1)}\leqslant 0,\,\forall n\geqslant 0$, is proved. 

\textit{Remark}.---Note that the choice $\zeta_{i}=0$ ($i=A,B$) may not cause any change in the modified total functional $\mathpzc{J}_{\,\Upsilon}$ with the change in the dynamics $\chi$. This means we have $\Delta\mathpzc{M}_{\,\Upsilon}=0,\;\forall\Delta\chi(t_{\mathrm{f}})$ and $\Delta\mathpzc{R}_{\,\Upsilon}=0\;\forall\Delta\chi(t),\, t\in[0,t_{\mathrm{f}}]$. Thus, by this choice in the worst case, the total objective functional $\mathpzc{J}_{\,\Upsilon}$ remains unchanged during the first step of the Krotov method. However, after minimizing the modified total functional $\mathpzc{J}_{\,\Upsilon}$ with respect to the field in the second step, the strict monotonic convergence almost always is preserved even by the choice $\zeta_{i}=0$ for one or both of $i=A$ and $i=B$.
\twocolumngrid
\end{widetext}

\bibliography{myreferences-resub}


\end{document}